\documentclass[usenatbib,useAMSm a4paper, singlespacing]{mn2e}

\usepackage{subcaption}
\usepackage{caption}
\usepackage{graphicx}	
\usepackage{amsmath}	
\usepackage{amssymb}	
\usepackage{bm}		
\usepackage{pdflscape}
\usepackage[T1]{fontenc}
\usepackage{txfonts}

\usepackage{float}

\usepackage{setspace}

\usepackage{lipsum}   
\usepackage{afterpage}

\usepackage{longtable,lscape, booktabs,etoolbox}
\usepackage{rotating}
\usepackage{multirow,multicol}

\usepackage{epsfig}
\usepackage{footnote}
\usepackage[referable]{threeparttablex}
\usepackage[normalem]{ulem}
\usepackage{color}
\usepackage{url}
\usepackage[breaklinks]{hyperref}
\usepackage{dblfloatfix}
\usepackage{placeins}
\usepackage{booktabs}

\def\mearth{\ifmmode {\rm M_{\oplus}}\else $\rm M_{\oplus}$\fi}
\def\Mearth{\ifmmode {\rm M_{\oplus}}\else $\rm M_{\oplus}$\fi}
\def\Rearth{\ifmmode {\rm R_{\oplus}}\else $\rm R_{\oplus}$\fi}
\def\Ms{\ifmmode {M_s}\else $M_s$\fi}
\def\Mp{\ifmmode {M_p}\else $M_p$\fi}
\def\Rp{\ifmmode {R_p}\else $R_p$\fi}
\def\rearth{\ifmmode {\rm R_{\oplus}}\else $\rm R_{\oplus}$\fi}

\def\aap{A\&A}

\def\apjl{ApJL}
\def\apjs{ApJS}

\newcommand{\Msun}{M_{\odot}}

\newcommand{\Lsun}{L_{\odot}}

\usepackage{booktabs,caption}

\title[ALMA SURVEY OF CIRCUMSTELLAR DISKS IN  IC 348]{ALMA SURVEY OF CIRCUMSTELLAR DISKS IN THE YOUNG STELLAR CLUSTER IC 348}

\author[D. Ru\'iz-Rodr\'iguez et al.]
{ \Large D. Ru\'iz-Rodr\'iguez, $^{1, 2}$\thanks{E-mail:dary.ruiz@anu.edu.au}
L. A. Cieza,$^{3,4}$
J. P. Williams,$^5$
S. M. Andrews,$^6$ 
D. A. Principe,$^{7}$
C. Caceres,$^{8, 12}$
\newauthor \Large H. Canovas,$^{9}$ 
S. Casassus,$^{3,10}$ 
M. R. Schreiber,$^{3, 11,12}$ and J. H. Kastner$^{1}$
\\
$^{1}$Chester F. Carlson Center for Imaging Science, School of Physics $\&$ Astronomy, and Laboratory for Multiwavelength Astrophysics,\\
Rochester Institute of Technology, 54 Lomb Memorial Drive, Rochester NY 14623 USA.\\
$^{2}$Research School of Astronomy and Astrophysics, Australian National University, Canberra, ACT 2611, Australia\\
$^{3}$Millenium Nucleus ``Protoplanetary discs in ALMA Early Science", Chile. \\
$^{4}$N\'ucleo de Astronom\'ia,  Facultad de Ingenier{\'i}a, Universidad Diego Portales,  Av. Ejercito 441, Santiago, Chile.\\
$^{5}$Institute for Astronomy, University of Hawaii at Manoa, Honolulu, HI, 96822, USA.\\
$^{6}$Harvard-Smithsonian Center for Astrophysics, 60 Garden Street, Cambridge, MA, 02138, USA.\\
$^{7}$Department of Physics and Kavli Institute for Astrophysics and Space Research, Massachusetts Institute of Technology, Cambridge, MA 02139, USA.\\
$^{8}$Departamento de Ciencias Fisicas, Facultad de Ciencias Exactas, Universidad Andres Bello. Av. Fernandez Concha 700, Las Condes, Santiago, Chile.\\
$^{9}$European Space Astronomy Centre (ESA), Camino Bajo del Castillo s/n, 28692 Villanueva de la Ca\~nada, Madrid, Spain.\\
$^{10}$Universidad de Chile, Camino el Observatorio 1515, Santiago, Chile.\\
$^{11}$Instituto de Física y Astronomía, Universidad de Valparaíso, Av. Gran Bretaña 1111, 2360102 Valparaíso, Chile.\\
$^{12}$N\'ucleo Milenio Formaci\'on Planetaria - NPF, Universidad de Valparaiso, Av. Gran Breta\~na 1111, Valparaiso, Chile.\\
}

\begin{document}

\date{}

\pagerange{\pageref{firstpage}--\pageref{lastpage}} \pubyear{2018}

\maketitle

\label{firstpage}

\begin{abstract}


We present  a 1.3 mm continuum survey of the young (2-3 Myr) stellar cluster IC~348, which lies at a distance of 310 pc, and is dominated by low-mass stars (M$_{\star}$ $\sim$ 0.1-0.6 M$_{\odot}$).  We observed 136 Class II sources (disks that are optically thick in the infrared) at 0.8$''$ (200 au) resolution with a 3$\sigma$ sensitivity of $\sim$ 0.45 mJy (M$_{\rm dust}$  $\sim$ 1.3 M$_{\oplus}$). 
We detect 40 of the targets and construct a mm-continuum luminosity function. We compare the disk mass  distribution  in IC 348  to those of younger and older regions, taking into account the dependence on stellar mass.  We find a clear evolution in disk masses from 1 to 5-10 Myr. The disk masses in IC 348 are significantly lower than those in Taurus (1-3 Myr) and Lupus (1-3 Myr), similar to those of  Chamaleon~I, (2-3 Myr)  and $\sigma$ Ori (3-5 Myr) and significantly higher than in Upper Scorpius (5$-$10 Myr).
About 20 disks in our sample ($\sim$5$\%$ of the cluster members) have estimated masses (dust $+$ gas) $>$1 M$_{\rm Jup}$ and hence might be the precursors of giant planets in the cluster. 
Some of the most massive disks include transition objects with inner opacity holes based on their infrared SEDs.  
From a stacking analysis of the 96 non-detections, we find that  these disks have a typical dust mass of just $\lesssim$ 0.4 M$_{\oplus}$, even though the vast majority of  their infrared SEDs remain optically thick and show little signs of evolution.  Such low-mass disks may be the precursors of the small rocky planets found by \emph{Kepler} around M-type stars. 
\end{abstract}

\begin{keywords}
Circumstellar Disks, Dust and Gas, Interferometry.
\end{keywords}

\section{Introduction}
\label{sec: intro}

The evolution of protoplanetary disks has been studied for decades, and typical disk lifetimes are well established to be $\sim$2-3 Myr \citep{WilliamsCieza2011}. 
On this timescale, the dust and gas components undergo significant evolution which, together with the initial conditions,  determine the outcome of the planet formation process.
By an age of $\sim$5 Myr, around 90$\%$ of protoplanetary disks  have already dispersed, constraining the time available for most planets to be formed \citep{SiciliaAguilar2006}. 
Determining the main process of disk dispersal is not an easy task since several physical mechanisms play a role at different time scales and radii \citep{Alexander2014}, but 
studying disk properties  as a function of stellar mass and age can shed light on the frequency and location of forming planets \citep{Mordasini2012, Alibert2011}. 

One important inference from exoplanet surveys is that planet occurrence generally decreases with increasing planet size:  rocky planets are much more common than gas giants  \citep{Howard2012, Burke2015, Bonfils2013}. 
Moreover, the correlation between stellar and planet properties indicates that giant planet occurrence increases with stellar mass at solar metallicity, with a percentage of 3$\%$ around M dwarfs ($\sim$0.5 M$_{\odot}$) increasing to 14$\%$ around A stars ($\sim$2 M$_{\odot}$) \citep{Johnson2010}. 
These exoplanet correlations are likely to be connected to disk properties as functions of stellar mass.

Emission from millimeter-sized dust grains in the disk is generally optically thin in the (sub-)millimeter regime; therefore, (sub-)millimeter continuum surveys of disks in star-forming regions with different ages ($\sim$1$-$10 Myr) can trace the distribution of disk masses as a function of age and stellar mass. This allows us to investigate how disk properties and evolution connect to the population of planets observed in the field. To exploit this observational potential, \citet{Andrews2013} performed a millimeter continuum survey with the Submillimeter Array (SMA) of the Taurus Class~II members (optically thick disks) with spectral types earlier than M8.5. As a main result, they showed a correlation between the mm luminosity (L$\rm _{mm}$) and the mass of the host stellar object of the form  L$\rm _{mm}$  $\propto$ M$_{\ast}$$^{1.5-2.0}$, which in turn suggests a linear relationship between the masses of the disk and that of the parent star:  M$_{\rm dust}$  $\propto$ M$_{\ast}$.

Various observational studies of higher sensitivity and resolution with the Atacama Large Millimeter/submillimeter Array (ALMA) add additional samples in Lupus \citep[1 $-$ 3 Myr;][]{Comeron2008, Alcala2014, Ansdell2016}, Chamaeleon I \citep[2 $-$ 3 Myr;][]{Luhman2007, Pascucci2016}, $\sigma$ Ori \citep[3 $-$ 5 Myr;][]{Oliveira2002, Ansdell2016}, and the Upper Scorpius OB Association \citep[5 $-$ 10 Myr;][]{Pecaut2012, Barenfeld2016}. 
A Bayesian linear regression has been the standard method used to characterize the M$\rm _{dust}$ - M$\rm _{\ast}$ relations of these star-forming regions. Although initially M$\rm _{dust}$ and M$\rm _{\ast}$ were thought to be linearly correlated  in 1$-$3 Myr old clusters, the main caveat of these linear relations is the simultaneous fitting of detections and upper limits, which is complicated by the fact that the latter dominate the aforementioned surveys. This adds more uncertainty to the M$\rm _{dust}$ - M$\rm _{\ast}$ relationship because the limited sensitivity implies lower detection rates for late-type stars and brown dwarfs, allowing for the possibility of a steeper relation. Indeed, \citet{Pascucci2016} reanalyzed all the submillimeter fluxes and stellar properties available for Taurus, Lupus, and Upper Sco, and found steeper correlations than linear for these clusters. They also obtained a steep dust mass-stellar mass scaling relation in the $\sim$ 2 Myr Cha I star-forming region, hence concluding that the same M$\rm _{dust}$ - M$\rm _{\ast}$ relation is shared by star-forming regions that are 1-3 Myr old \citep{Pascucci2016}. More recently, a similar steepening of the M$\rm _{dust}$ - M$\rm _{\ast}$ relation was found by \citet{Ansdell2017} for the $\sigma$ Ori star-forming region. This steeper relation possibly indicates 1) undetected large pebbles or 2) an efficient inward drift in disks around the lowest-mass stars. In addition, this steepening of the M$\rm _{dust}$ - M$\rm _{\ast}$ correlation with age suggests a faster decline of circumstellar dust mass with time in late-type stars. From these relations, at an age of $\la$10 Myrs, disks around 0.1 and 0.5 M$_{\odot}$ stars might have dispersed millimeter-sized grains by factors of 5 and 2.5, respectively, faster than earlier-type objects \citep{Pascucci2016}.

Following these studies, the IC 348 star-forming region, with a disk fraction of 36$\%$ in the IR regime, is an excellent benchmark to characterize the relationship between the masses of the disk and that of the host star by comparing to other star-forming regions. In fact, the first millimeter observations of protoplanetary disks in IC348 star-forming region were made by \citet{Lee2011}, and with a detection rate of only $\sim$ 12$\%$, they concluded that most of the solids in the IR-detected disks have aggregated beyond millimeter sizes, resulting in low luminosities at millimeter wavelengths.

In this work, we present a 1.3 mm/230 GHz study of $\sim$136 Class~II objects in the IC 348 star-forming region. This paper is organised as follows: Section \ref{Sec:Target} describes the target selection.  Section \ref{Sec:obs_ic348} summarizes the ALMA observations and data reduction. In Section \ref{Sec:Results}, we estimate the stellar properties of our sample, and present our ALMA results,  which are compared to previous  findings in other regions in Section \ref{Sec:Discussion}. 
The main conclusions are discussed in Section \ref{Sec:Summary}.

\begin{figure}
 \centering
  \includegraphics[width=0.51\textwidth]{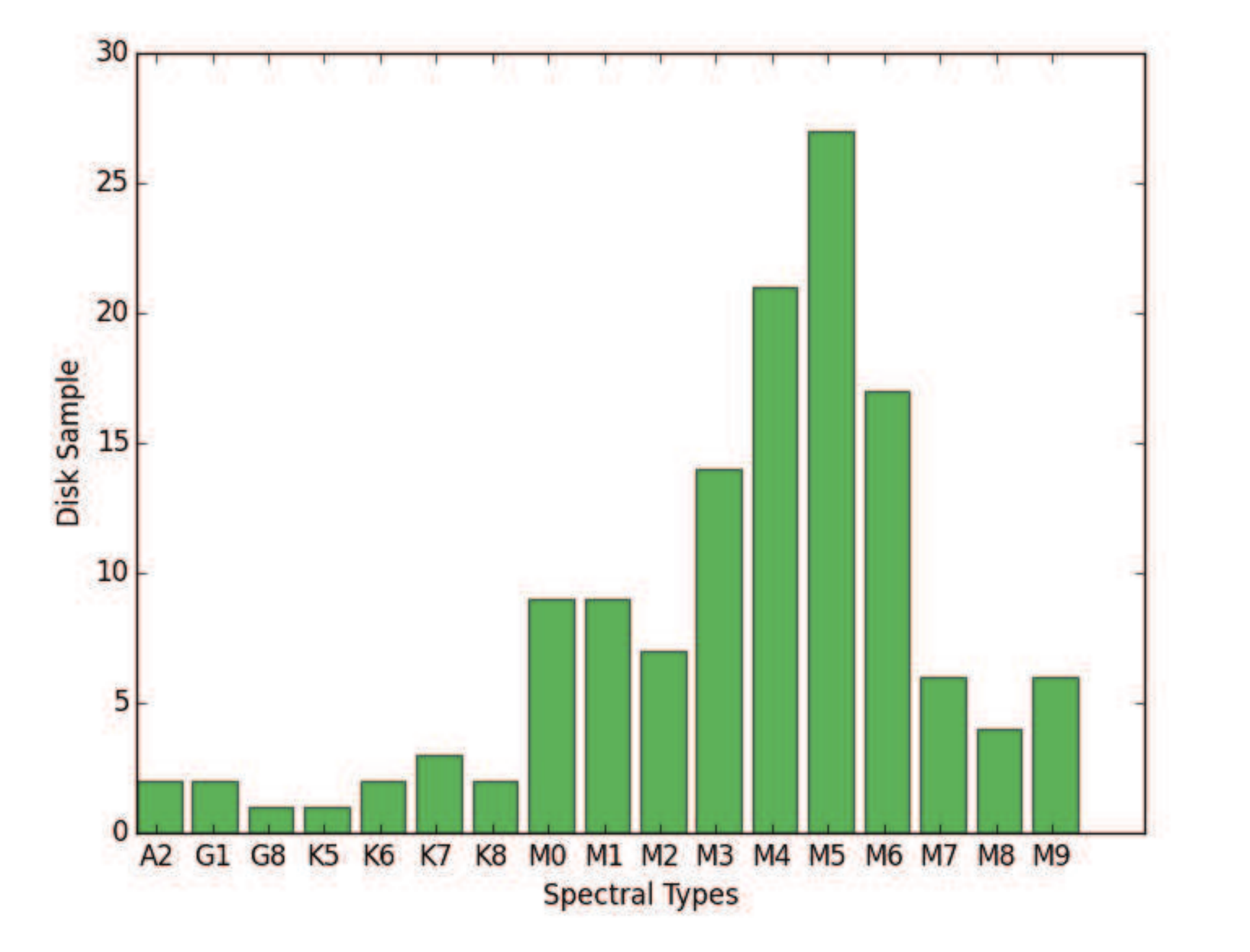}
   \caption[Distribution of Stellar Spectral Types in IC 348]{Distribution of stellar spectral types for our sample in the IC 348 star-forming region. These targets were selected from \citet{Muench2007} and \citet{Lada2006}
   and  are listed in Table \ref{Table:sample}}
  \label{Fig:distribution}
\end{figure}

\section{TARGET SELECTION AND PROPERTIES}
\label{Sec:Target}

IC~348  is a rich and compact (2$\times$2 pc) young stellar cluster in the Perseus molecular cloud,
 
whose $\sim$480 members have been identified initially by H$\alpha$ emission \citep{Herbig1954} and subsequently by optical and IR photometry and spectroscopy \citep{Lada1995, Herbig1998, Luhman1998, Luhman1999, Luhman2003, Luhman2003b, Luhman1999, Luhman2005, Luhman2016}. Most of the known T Tauri stars in the IC348 star-forming region have been well studied and spectrally classified \citep{Luhman2003, Muench2007}: see Figure \ref{Fig:distribution}.

Our sample was selected specifically from the work of \citet{Lada2006}, whose sample was based on \citet{Luhman2003}, and from \citet{Muench2007}, a census of 192 candidate YSOs in the IC 348 nebula, covering a 26.8 $^{'}$$\times$ 28.5 $^{'}$ region and centered at R.A. 03$^{h}$44$^{m}$20.52$^{s}$, Dec. $+$32$^{o}$10$^{'}$34.87$^{''}$. These programs used \textit{Spitzer}-IRAC photometry to investigate both the frequency and nature of the circumstellar disk population in the IC348 cluster on the basis of the IR SED slope between 3.6 and 8.0 $\mu$m,  $\alpha _{3.6-8.0\mu m}$. 
In general, \citet{Lada2006} and \citet{Muench2007} used $\alpha _{3.6-8.0\mu m}$ to classify the objects as follows:

\begin{enumerate}
\item Class I (protostars):  $\alpha _{3.6-8.0\mu m}$ $>-$0.5;
\item Class II (thick-disks):  $-$0.5$>$ $\alpha _{3.6-8.0\mu m}$ $>$ $-$1.8;
\item Class II/III (anemic disks): $-$1.8 $>$ $\alpha _{3.6-8.0\mu m}$ $>$ $-$2.56;
\item Class III (disk-less stars): $\alpha _{3.6-8.0\mu m}$ $<-$2.56.
\end{enumerate}

\begin{figure}
 \centering
  \includegraphics[width=0.45\textwidth]{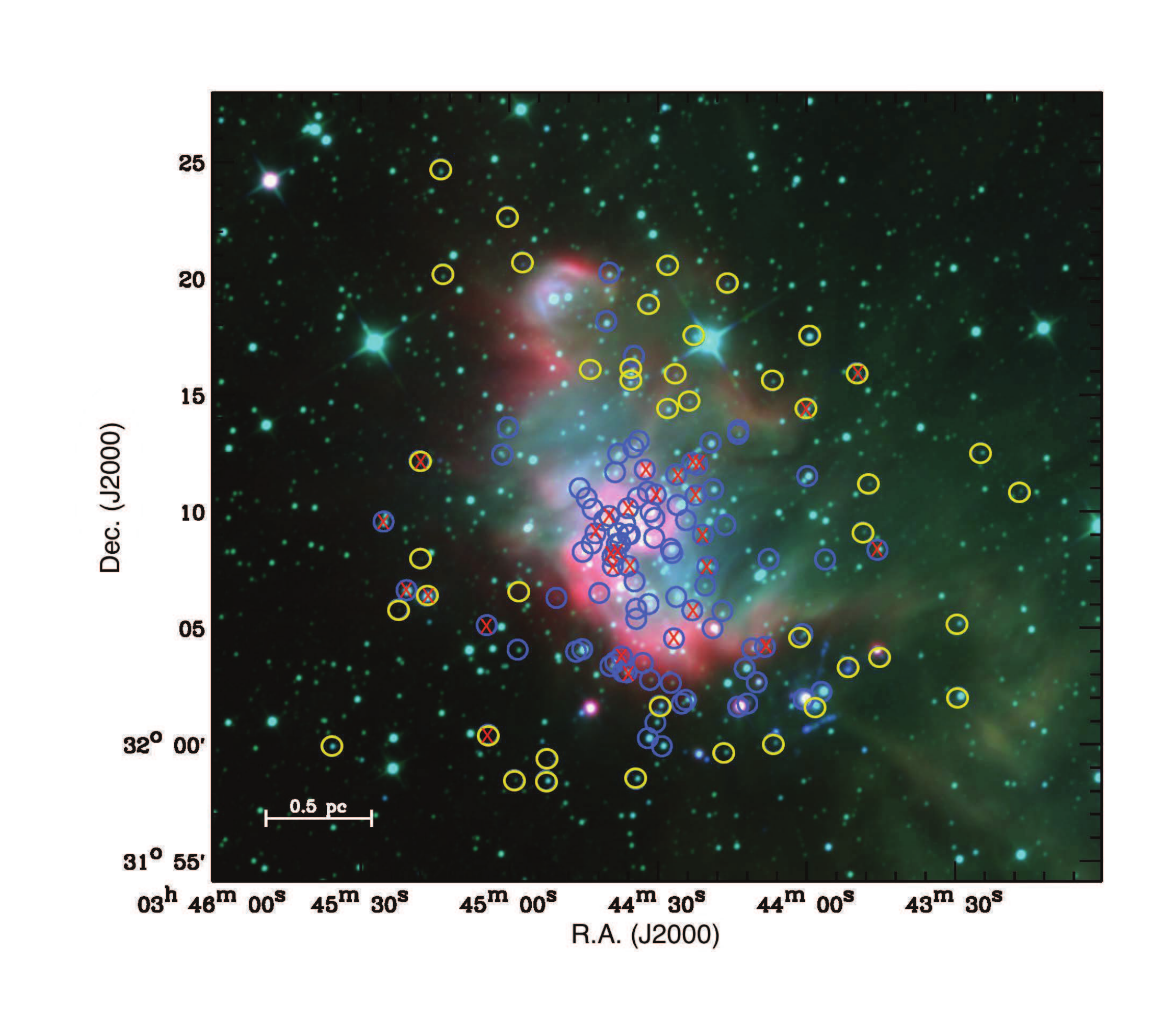}      
   \caption[IR map of IC 348 star-forming region]{IR map of IC 348 star-forming region with our ALMA targets. ALLWISE 3-color image with RGB mapped to 22 (W4), 4.6 (W2), and 3.4 (W1) $\micron$. Yellow and blue circles correspond to the sampled selected from \citet{Muench2007} and \citet{Lada2006}, respectively. Red crosses indicate the positions of IC~348 members used to estimate a mean cluster distance of 310 pc $\pm$ 20 pc, based on the ${\it Gaia}$ DR2 parallax data.}
  \label{Fig:area}
\end{figure}

\citet{Lada2006} classified  as ``anemic disks'' those objects with $-$1.8 $>$ $\alpha _{3.6-8.0\mu m}$ $>$ $-$2.56, while  \citet{Muench2007} did not search for members with ``anemic'' type disks. Hence, we selected \textit{Spitzer} sources with  $\alpha _{3.6-8.0\mu m}$ values between $-$1.8 and $-$0.5, which corresponds to Class~II T Tauri stars with optically thick disks. From \citet{Lada2006}, we selected 91 objects classified as optically ``THICK'' disks (hereafter Class II sources to keep the nomenclature consistent), and from \citet{Muench2007}, we selected 42 objects classified as Class~II objects. We also included Cl* IC  348  LRL  31, Cl* IC  348  LRL 67 and Cl* IC  348  LRL  329,  which are Class III sources based on their $\alpha _{3.6-8.0\mu m}$ values, but their 24 $\mu$m fluxes indicate that they are transitional objects with optically thick outer disks \citep{Lada2006, Espaillat2012}. 
We note that the standard YSO Class system \citep{Greene1994} is based on the SED slope between $\sim$2 and $\sim$20 $\mu$m, but most IC 348 members lack \emph{Spitzer} 24 $\mu$m detections. 
With these caveats, our final target list (Table \ref{Table:sample}) is composed of 136 Class II disk objects with stellar spectral types in the range of G1$-$M9. Figure \ref{Fig:area} shows the positions of our targets. Among the objects selected, Cl* IC  348  LRL  237, V* V716 Per,  Cl* IC  348  LRL  135 and Cl* IC  348  LRL  97 are classified by \citet{Espaillat2012} as transitional disks. Our sample also includes Cl* IC  348  LRL  31 and Cl* IC  348  LRL  135, which have known close stellar companions at separations of 38.1$\pm$ 5.3 mas and 82.1$\pm$ 0.3 mas, respectively \citep{RuizRodriguez2016}.

\section{ALMA Observations and Data Reduction}
\label{Sec:obs_ic348}

ALMA observations toward our IC 348 targets were carried out in Band 6 (211-275 GHz) under the project code: 2015.1.01037.S. Our science goal was executed in Cycle 3 with the C40-4 array configuration and was observed between the 23rd and 27th June, 2015. The Band 6 continuum observations were conducted with a total on-source integration time of $\sim$1 min per target over 3 execution blocks,  each one targeting all 136 objects for 0.3 min. The adopted setup included two spectral windows for continuum observations with effective bandwidths of 1.875  GHz centered at  218.0 and 233.0 GHz,  for a mean frequency of 225.676 GHz ($\sim$1.3 mm). The typical (1$\sigma$) noise level reached is $\sim$0.15 mJy/beam. 
 We also targeted molecular emission lines of $^{12}$CO, $^{13}$CO,  and C$^{18}$O (J = 2$-$1), centered on 230.535, 220.395, 219.557 GHz, respectively. Each line was observed with a resolution of 242 kHz (0.3 km s$^{-1}$) and a bandwidth of 117.2 MHz. 
The ALMA data were reduced using the Common Astronomy Software Application (CASA) package, version 4.5.3 \citep{McMullin2007}. Initial calibration (i.e. water vapor radiometer corrections,  phase and amplitude calibrations)  was performed by the ALMA science operations team during quality assurance. The flux calibrator was J0237+2848.  J0238+1636 and J0336+3218 were chosen as bandpass calibrators and J0510+1800 as a phase calibrator. To reach the requested synthesized beam size of $\sim$0.8 arcsec, we applied the CLEAN algorithm to the calibrated visibilities and extracted the continuum images by applying a Briggs weighting with a robust parameter of +2, which is close to a natural weighting. Using the \textit{uvcontsub} routine, we subtracted the continuum emission from the spectral windows to extract the $^{12}$CO, $^{13}$CO, and C$^{18}$O spectral line data from the calibrated visibilities. 

\vspace{-0.1cm}

\section{Results}
\label{Sec:Results}

\subsection{Detection Criteria}
\label{Sec:Criteria}

We searched for 1.3 mm continuum emission centered on the 2MASS positions of the 136 targets, listed in Table \ref{Table:sample}. From the continuum images, we determined the peak flux and rms using the task \textit{imstat} and thus estimated the signal to noise  (S/N, ratio between peak and rms) for each image. Peak fluxes were derived from a 4$^{''}$ radius circle, and the rms from a 4-7$^{''}$ radius annulus centered on the expected source position. A source with S/N $<$ 4 is considered a non-detection. For these sources, we measured the flux densities by using the \textit{uvmodelfit} routine in CASA and by fitting a point source in the \textit{uv} plane. If the flux density is less than 4$\sigma$, the point source fit is applied to the visibilities with the pointing center as a free parameter. If the flux density is less than 3$\sigma$ it is fit with a point source with the offset position fixed. Table \ref{Table:observations_no} lists integrated flux density ($\rm F_{1.3 \, mm}$) and rms for non-detected sources.

\begin{figure*}
 \centering
  \includegraphics[width=0.83\textwidth]{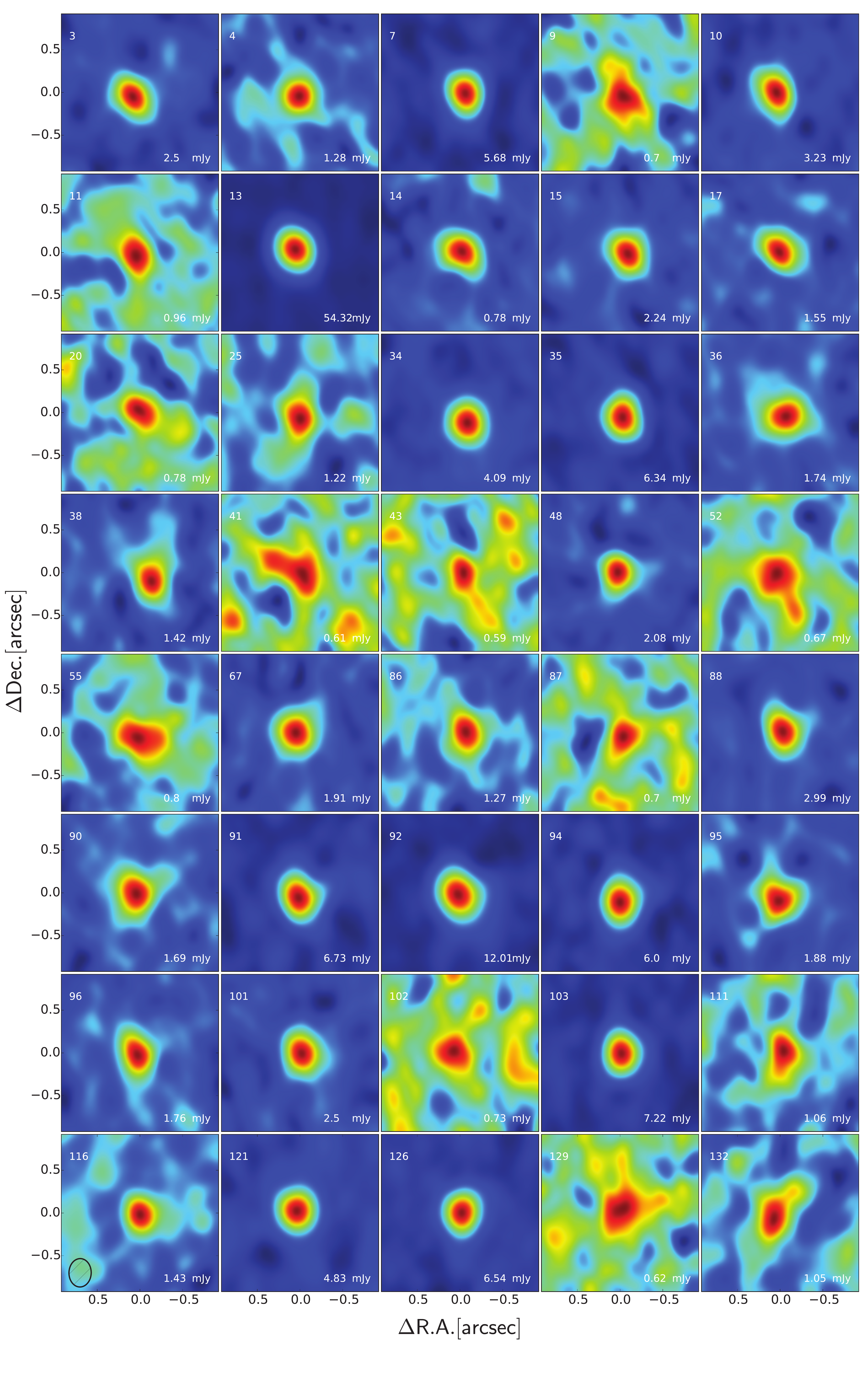}
            
   \caption[1.3 mm continuum images of IC 348]{1.3 mm continuum images meeting our detection criteria  ($>$4$\sigma$) in the IC 348 region, see Section \ref{Sec:Criteria}. Each image covers 1.7'' X 1.7'' size with an average beam size of 0.8$^{''}$ X 0.7$^{''}$. Integrated flux density values are presented at the low-right corner as reported in Table \ref{Table:observations}.}
  \label{fig:continuum}
\end{figure*}

For detections (S/N $>$ 4), flux densities were measured by applying an elliptical Gaussian model to the visibility  data using \textit{uvmodelfit} in CASA. This model is centered at the nominal source position and provides the parameters $\rm F_{1.3 \, mm}$, the FWHM along the major axis, aspect ratio, position angle of the major axis (P.A.), and  coordinate offsets ($\rm \Delta \alpha$, and $\rm \Delta \delta$). These parameters are listed in Table \ref{Table:observations}. A disadvantage of fitting the brightness profile of a source in the UV-plane directly is the possibility of including emission from a second source in the fitting process. To avoid any contamination in the measured flux of each field, we visually inspected the image plane for pixels with significant brightness ($>$4$\sigma$). Applying these methods, we detect 40 out of the 136 IC 348 targets at $>$~4$~\sigma$ significance. Images of the 40 sources are displayed in Figure \ref{fig:continuum}. We find that 10 of the targets are partially resolved, giving P.A. values with large uncertainties, and therefore, we do not report those values here. For these objects, the source sizes  (deconvolved from the beam) are listed in Table \ref{Table:observations}.

Using standard approaches \citep[e.g.][]{Hildebrand1983}, the millimeter flux can be translated into a disk mass according

\begin{equation}
M_{dust}  =  \frac{F_\nu d^2}{\kappa_\nu B_\nu (T_{dust})},
\end{equation}

\newpage
\onecolumn
\renewcommand{\thefootnote}{\fnsymbol{footnote}}
\renewcommand{\arraystretch}{0.81}
\begin{longtable}{cccccl}
\caption[Targeted Class II Objects in IC~348 ]{Targeted Class II Objects in IC~348.}
\label{Table:sample} \\
\hline \hline \\[-0.9ex]
   \multicolumn{1}{c}{\textbf{Source ID}} &
   \multicolumn{1}{c}{\textbf{Target}} &
   \multicolumn{1}{c}{\textbf{R.A.}} &
   \multicolumn{1}{c}{\textbf{Dec.}} &
    \multicolumn{1}{c}{\textbf{Spec. type}}  &
    \multicolumn{1}{c}{\textbf{Ref.}} \\[-2.0ex] 
    \multicolumn{1}{c}{\textbf{}} &
   \multicolumn{1}{c}{\textbf{}} &
   \multicolumn{1}{c}{\textbf{}} &
  \multicolumn{1}{c}{\textbf{}} &
   \multicolumn{1}{c}{\textbf{}} &
     \multicolumn{1}{c}{\textbf{}} \\[1.5ex]\hline \hline \\[-1ex]
\endfirsthead

\multicolumn{6}{c}{{\tablename} \thetable{} -- Continued}\\[1.0ex]
  \hline \hline \\[-0.9ex]
   \multicolumn{1}{c}{\textbf{Source}} &
   \multicolumn{1}{c}{\textbf{Target}} &
   \multicolumn{1}{c}{\textbf{R.A.}} &
   \multicolumn{1}{c}{\textbf{Dec.}} &
   \multicolumn{1}{c}{\textbf{Spect. type}}  &
   \multicolumn{1}{c}{\textbf{Ref.}} \\[-2.0ex] 
    \multicolumn{1}{c}{\textbf{}} &
   \multicolumn{1}{c}{\textbf{}} &
   \multicolumn{1}{c}{\textbf{}} &
  \multicolumn{1}{c}{\textbf{}} &
   \multicolumn{1}{c}{\textbf{}} &
     \multicolumn{1}{c}{\textbf{}} \\[1.5ex]\hline \hline \\[-1ex]
   \endhead
\endfoot

\hline\hline\\[-1.4ex] 
\endlastfoot

1	&	IC  348   12	&	03 44 35.34	&	+32 10 04.88	&	A2		&	1	\\
2	&	V* V909 Per	&	03 44 26.03  	&	+32 04 30.41  	&	G8		&	1	\\
3	&	Cl* IC  348  LRL  13	&03 43 59.65  	&	+32 01 53.98  	&	M0.5	&	1	\\
4	&	V* V926 Per	&	03 44 44.72  	&	+32 04 02.48  	&	M0.5		&	1	\\
5	&	Cl* IC  348  LRL  19	&	03 44 30.82   	&	+32 09 55.80  	&	A2		&	1	\\
6	&	Cl* IC  348  LRL   26	&	03 43 56.03  	&	+32 02 13.21  	&	K7     	&	1	\\
7	&	V* V920 Per	&	03 44 37.88  	&	+32 08 04.18  	&	K7		&	1	\\
8	&	V* V715 Per	&	03 44 38.46  	&	+32 07 35.70  	&	K6		&	1	\\
9	&	V* V712 Per	&	03 44 37.98  	&	+32 03 29.66  	&	K6		&	1	\\
10	&	V* V910 Per	&	03 44 29.73  	&	+32 10 39.84  	&	K8		&	1	\\
11	&	V* V697 Per	&	03 44 21.61  	&	+32 10 37.68  	&	K7		&	1	\\
12	&	Cl* IC  348  LRL  46	&	03 44 11.62  	&	+32 03 13.18  	&  --	&	1	\\
13	&	IRAS 03410+3152	&	03 44 12.98  	&	+32 01 35.50  	&	--		&	1	\\
14	&	Cl* IC  348  LRL  55	&	03 44 31.37  	&	+32 00 14.05  	&	M0.5		&	1	\\
15*	&	V* V716 Per	&	03 44 38.54  	&	+32 08 00.65  	&	M1.25	&	1, 3	\\
16	&	V* V698 Per	&	03 44 22.29  	&	+32 05 42.79  	&	K8		&	1	\\
17	&	Cl* IC  348  LRL  63	&	03 43 58.91  	&	+32 11 27.07  	&	M1.75		&	1	\\
18	&	Cl* IC  348  LRL  68	&	03 44 28.51  	&	+31 59 54.00  	&	M3.5		&	1	\\
19	&	V* V719 Per	&	03 44 43.77  	&	+32 10 30.41  	&	M1.25		&	1	\\
20	&	Cl* IC  348  LRL  76	&	03 44 39.80  	&	+32 18 04.19  	&	M3.75		&	1	\\
21	&	V* V710 Per	&	03 44 37.41  	&	+32 09 00.91  	&	M1		&	1	\\
22	&	V* V922 Per	&	03 44 39.20 	&	+32 09 44.90  	&	M2		&	1	\\
23*	&	Cl* IC  348   LRL  97	&	03 44 25.55  	&	+32 06 17.13  	&	M2.25		&	1,3	\\
24	&	V* V695 Per	&	03 44 19.24  	&	+32 07 34.74  	&	M3.75		&	1	\\
25	&	V* V905 Per	&	03 44 22.32  	&	+32 12 00.70  	&	M1	&		1	\\
26	&	V* V925 Per	&	03 44 44.59  	&	+32 08 12.54  	&	M2	&		1	\\
27	&	V* V919 Per	&	03 44 37.39  	&	+32 12 24.20  	&	M2	&		1	\\
28	&	Cl* IC  348  LRL  128	&	03 44 20.18  	&	+32 08 56.59  	&	M2		&	1	\\
29	&	Cl* IC  348  LRL  129	&	03 44 21.30  	&	+32 11 56.34  	&	M2		&	1	\\
30*	& Cl* IC  348  LRL 135	&	03 44 39.18  	&	+32 20 08.93  	&	M4.5		&	1,3	\\
31	&	V* V907 Per	&	03 44 25.30  	&	+32 10 12.80  	&	M4.75		&	1	\\
32	&	Cl* IC  348  LRL  140	&	03 44 35.69  	&	+32 03 03.54  	&	M3.25		&	1	\\
33	&	Cl* IC  348  LRL  149	&	03 44 36.98   	&	+32 08 34.20   	&	M4.75		&	1	\\
34	&	Cl* IC  348  LRL  153	&	03 44 42.76  	&	+32 08 33.77  	&	M4.75		&	1	\\
35	&	Cl* IC  348  LRL  156	&	03 44 06.78  	&	+32 07 54.09  	&	M4.25		&	1	\\
36	&	V* V902 Per	&	03 44 18.58  	&	+32 12 53.08  	&	M2.75		&	1	\\
37	&	Cl* IC  348  LRL  165	&	03 44 35.46  	&	+32 08 56.35  	&	M5.25		&	1	\\
38	&	Cl* IC  348  LRL  166A	&	03 44 42.58  	&	+32 10 02.50  	&	M4.25		&	1	\\
39	&	Cl* IC  348  LRL  168	&	03 44 31.35  	&	+32 10 46.98  	&	M4.25		&	1	\\
40	&	Cl* IC  348  LRL  173	&	03 44 10.13  	&	+32 04 04.50  	&	M5.75		&	1	\\
41	&	Cl* IC  348  LRL  192	&	03 44 23.65  	&	+32 01 52.69  	&	M4.5		&	1	\\
42	&	V* V713 Per	&	03 44 38.01  	&	+32 11 37.03  	&	M4		&	1	\\
43	&	Cl* IC  348  LRL  202	&	03 44 34.28  	&	+32 12 40.73  	&	M3.5		&	1	\\
44	&	Cl* IC  348  LRL  203	&	03 44 18.10  	&	+32 10 53.44  	&	M0.75		&	1	\\
45	&	Cl* IC  348  LRL  205	&	03 44 29.80  	&	+32 00 54.58  	&	M6		&	1	\\
46	&	Cl* IC  348  LRL  214	&	03 44 07.51  	&	+32 04 08.81  	&	M4.75		&	1	\\
47	&	Cl* IC  348  LRL  221	&	03 44 40.24  	&	+32 09 33.13  	&	M4.5		&	1	\\
48	&	SSTc2d J034431.2+320559	&	03 44 31.19   	&	+32 05 58.90   	&	M0.5		&	1	\\
49*	&	Cl* IC  348    LRL  229	&	03 44 57.86  	&	+32 04 01.60  	&	M5.25		&	1, 3	\\
50	&	Cl* IC  348  LRL  237	&	03 44 23.57  	&	+32 09 33.88  	&	M5		&	1	\\
51	&	Cl* IC  348   LRL  241	&	03 44 59.83   	&	+32 13 31.90   	&	M4.5	&	1	\\
52	&	Cl* IC  348  LRL  248	&	03 44 35.95  	&	+32 09 24.31  	&	M5.25		&	1	\\
53	&	Cl* IC  348  LRL  256	&	03 43 55.27  	&	+32 07 53.31  	&	M5.75		&	1	\\
54	&	Cl* IC  348  LRL  272	&	03 44 34.13  	&	+32 16 35.77  	&	M4.25		&	1	\\
55	&	Cl* IC  348  LRL  276	&	03 44 09.21  	&	+32 02 37.68  	&	M0		&	1	\\
56	&	Cl* IC  348  H  149	&	03 44 34.05  	&	+32 06 57.05  	&	M7.25		&	1	\\
57	&	Cl* IC  348  LRL  292	&	03 43 59.87  	&	+32 04 41.44  	&	M5.75		&	1	\\
58	&	Cl* IC  348  LRL  297	&	03 44 33.21  	&	+32 12 57.46  	&	M4.5		&	1	\\
59	&	Cl* IC  348  LRL  300	&	03 44 38.97  	&	+32 03 19.69  	&	M5		&	1	\\
60	&	Cl* IC  348  LRL  319	&	03 45 01.00  	&	+32 12 22.21  	&	M5.5		&	1	\\
61	&	Cl* IC  348  LRL  324	&	03 44 45.22  	&	+32 10 55.75  	&	M5.75		&	1	\\
62	&	Cl* IC  348  LRL  325	&	03 44 30.06   	&	+32 08 48.90   	&	M6		&	1	\\
63	&	Cl* IC  348  LRL  334	&	03 44 26.66  	&	+32 02 36.32  	&	M5.75		&	1	\\
64	&	Cl* IC  348  LRL  336	&	03 44 32.37  	&	+32 03 27.48  	&	M5.5		&	1	\\
65	&	Cl* IC  348  LRL  341	&	03 44 12.98  	&	+32 13 15.61  	&	M5.25		&	1	\\
66	&	Cl* IC  348 LRL   366	&	03 44 35.02  	&	+32 08 57.34  	&	M4.75		&	1	\\
67	&	Cl* IC  348  LRL  382	&	03 44 30.96  	&	+32 02 44.18  	&	M5.5		&	1	\\
68	&	Cl* IC  348  LRL  407	&	03 45 04.14  	&	+32 05 04.38  	&	M7		&	1	\\
69	&	Cl* IC  348  LRL  415	&	03 44 29.97  	&	+32 09 39.45  	&	M6.5		&	1	\\
70	&	2MASS J03444593+3203567	&	03 44 45.94  	&	+32 03 56.78  	&	M5.75		&	1	\\
71	&	Cl* IC  348  LRL  462	&	03 44 24.46  	&	+32 01 43.71  	&	M3		&	1	\\
72	&	Cl* IC  348  LRL  468	&	03 44 11.07  	&	+32 01 43.60  	&	M8.25		&	1	\\
73	&	Cl* IC  348  LRL  555	&	03 44 41.22  	&	+32 06 27.14  	&	M5.75		&	1	\\
74	&	Cl* IC  348  LRL  603	&	03 44 33.42   	&	+32 10 31.50   	&	M8.5	&	1	\\
75	&	[PSZ2003] J034437.6+320832	&	03 44 37.64  	&	+32 08 32.90  	&	M5.5		&	1	\\
76	&	[PSZ2003] J034426.4+320809	&	03 44 26.37  	&	+32 08 09.94  	&	M9		&	1	\\
77	&	Cl* IC  348  LRL  690	&	03 44 36.38   	&	+32 03 05.40   	&	M8.75		&	1	\\
78	&	Cl* IC  348  LRL  703	&	03 44 36.62   	&	+32 03 44.20  	&	M8		&	1	\\
79	&	[PSZ2003] J034433.7+320521	&	03 44 33.70  	&	+32 05 20.67  	&	M6		&	1	\\
80	&	[PSZ2003] J034433.7+320547	&	03 44 33.69  	&	+32 05 46.71  	&	M8.75		&	1	\\
81	&	Cl* IC  348  LRL  746	&	03 44 49.96  	&	+32 06 14.61  	&	M5		&	1	\\
82	&	[PSZ2003] J034419.7+320645	&	03 44 19.67  	&	+32 06 45.93  	&	M7		&	1	\\
83	&	Cl* IC  348  LRL 2096	&	03 44 12.94  	&	+32 13 24.06  	&	M6		&	1	\\
84	&	[PSZ2003] J034416.2+320540	&	03 44 16.18  	&	+32 05 40.96  	&	M9		&	1	\\
85	&	Cl* IC  348  TJ  72	&	03 44 31.98  	&	+32 11 43.95  	&	G0		&	1	\\
86	&	[BNM2013] 32.03    53	&	03 44 42.01  	&	+32 08 59.98  	&	M4.25		&	1	\\
87	&	Cl* IC  348  LRL 8078	&	03 44 26.68  	&	+32 08 20.35  	&	M0.5		&	1	\\
88	&	Cl* IC  348  LRL 9024	&	03 44 35.37  	&	+32 07 36.24  	&	M0		&	1	\\
89	&	Cl* IC  348   H  110	&	03 44 25.58  	&	+32 11 30.24  	&	M2		&	1	\\
90	&	2MASS J03452514+3209301	&	03 45 25.15  	&	+32 09 30.18  	&	M3.75		&	1	\\
91	&	2MASS J03452046+3206344	&	03 45 20.46  	&	+32 06 34.48  	&	M1		&	1	\\
92*	&	Cl* IC  348  LRL  31&	03 44 18.17  	&	+32 04 57.04  	&	G1		&	1	\\
93*	&	Cl* IC  348  LRL  329	&	03 44 15.58  	&	+32 09 21.83  	&	M7.5		&	1	\\
94* &	Cl* IC  348  LRL   67		&	03 43 44.61  	&	+32 08 17.76  	&	M0.75		&	1	\\
95	&	2MASS J03435856+3217275	&	03 43 58.57  	&	+32 17 27.53  	&	M3.5(IR)		&	2	\\
96	&	Cl* IC  348  LRL  117	&	03 43 59.08  	&	+32 14 21.31  	&	M3.5(IR)		&	2	\\      
97	&	2MASS J03442724+3214209	&	03 44 27.25  	&	+32 14 20.98  	&	M3.5(IR)		&	2	\\
98	&	2MASS J03434881+3215515	&	03 43 48.81  	&	+32 15 51.55  	&	M4.5(IR)		&	2	\\
99	&	Cl* IC  348  LRL  179	&	03 44 34.99  	&	+32 15 31.15  	&	M3.5(IR)		&	2	\\
100	&	Cl* IC  348  LRL  199	&	03 43 57.22   	&	+32 01 33.90   	&	M6.75(IR)		&	2	\\
101	&	Cl* IC  348  LRL  215	&	03 44 28.95  	&	+32 01 37.85  	&	M3.25(IR)		&	2	\\
102	&	2MASS J03443112+3218484	&	03 44 31.13  	&	+32 18 48.49  	&	M3.25(IR)		&	2	\\
103*	&     2MASS J03443468+3216000	&	03 44 34.69  	&	+32 16 00.09  	&	M3.5(IR)		&	2, 3	\\
104	&	2MASS J03441522+3219421	&	03 44 15.22  	&	+32 19 42.18  	&	M4.75(IR)		&	2	\\
105	&	2MASS J03442294+3214404	&	03 44 22.94  	&	+32 14 40.43  	&	M5.5(IR)		&	2	\\
106	&	2MASS J03440599+3215321	&	03 44 05.99  	&	+32 15 32.15  	&	M6.5(IR)		&	2	\\
107	&	Cl* IC  348  LRL  364	&	03 44 43.01  	&	+32 15 59.67  	&	M4.75(IR)	&	2	\\
108	&	Cl* IC  348  LRL  368	&	03 44 25.70  	&	+32 15 49.27  	&	M5.5(IR)		&	2	\\
109	&	Cl* IC  348  LRL  406	&	03 43 46.44  	&	+32 11 05.94  	&	M5.75(IR)		&	2	\\
110	&	2MASS J03445853+3158270	&	03 44 58.54  	&	+31 58 27.03  	&	M6.5(IR)		&	2	\\
111	&	2MASS J03432845+3205058	&	03 43 28.45  	&	+32 05 05.82  	&	M4(IR)		&	2	\\
112	&	Cl* IC  348  LRL  753	&	03 44 57.62  	&	+32 06 31.25  	&	XXX		&	2	\\
113	&	2MASS J03445688+3220355	&	03 44 56.88  	&	+32 20 35.52  	&	M6(IR)		&	2	\\
114	&	Cl* IC  348  LRL 1379	&	03 44 52.00   	&	+31 59 21.50   	&	M9.75		&	2	\\
115	&	2MASS J03445205+3158252	&	03 44 52.06  	&	+31 58 25.21  	&	M3.5(IR)		&	2	\\
116	&	Cl* IC  348  LRL 1683	&	03 44 15.83  	&	+31 59 36.77  	&	M5.25(IR)		&	2	\\
117	&	Cl* IC  348  LRL 1707	&	03 43 47.64  	&	+32 09 02.56  	&	M7(IR)		&	2	\\
118	&	2MASS J03451307+3220053	&	03 45 13.07  	&	+32 20 05.32  	&	M5(IR)		&	2	\\
119	&	2MASS J03442721+3220288	&	03 44 27.21  	&	+32 20 28.82  	&	M5(IR)		&	2	\\
120	&	2MASS J03435056+3203180	&	03 43 50.57  	&	+32 03 18.00  	&	M8.75(IR)		&	2	\\
121	&	Cl* IC  348  LRL 1881	&	03 44 33.79  	&	+31 58 30.28  	&	M3.75(IR)		&	2	\\
122	&	2MASS J03432355+3212258	&	03 43 23.56  	&	+32 12 25.82  	&	M4.5(op)		&	2	\\
123	&	V* V338 Per	&	03 43 28.20  	&	+32 01 59.12  	&	M1.75(IR)		&	2	\\
124	&	Cl* IC  348  LRL 1923	&	03 44 00.47  	&	+32 04 32.71  	&	M5(IR)		&	2	\\
125	&	Cl* IC  348  LRL 1925	&	03 44 05.77  	&	+32 00 01.10  	&	M5.5(IR)		&	2	\\
126	&	EM* LkHA   99	&	03 45 16.35  	&	+32 06 19.95  	&	K5(op)		&	2	\\
127	&	2MASS J03445997+3222328	&	03 44 59.98  	&	+32 22 32.83  	&	M5.25		&	2	\\
128	&	2MASS J03451782+3212058	&	03 45 17.83  	&	+32 12 05.85  	&	M3.75(op)	&	2	\\
129	&	2MASS J03431581+3210455	&	03 43 15.81  	&	+32 10 45.53  	&	M4.5(IR)		&	2	\\
130	&	2MASS J03453563+3159544	&	03 45 35.64  	&	+31 59 54.44  	&	M4.5(IR)		&	2	\\
131	&	2MASS J03452212+3205450	&	03 45 22.13  	&	+32 05 45.01  	&	M8(IR)		&	2	\\
132	&	2MASS J03442186+3217273	&	03 44 21.86  	&	+32 17 27.31  	&	M4.75(op)		&	2	\\
133	&	Cl* IC  348  LRL  22865	&	03 45 17.65  	&	+32 07 55.33  	&	L0		&	2	\\
134	&	Cl* IC  348  LRL  40182	&	03 45 03.83   	&	+32 00 23.30   	&  --		&	2	\\
135	&	Cl* IC  348  LRL  54299	&	03 43 44.27   	&	+32 03 42.60   	&  --		&	2	\\
136	&	2MASS J03451349+3224347	&	03 45 13.50 	&	+32 24 34.71  	&	M4.25		&	2\\
\end{longtable}
\footnotesize{Ref.: (1) \citet{Lada2006}, (2) \citet{Muench2007}, (3) \citet{Espaillat2012}.}\\
\footnotesize{$*$ Transitional Disk }
\normalsize

\twocolumn

where $F_\nu$ is the integrated flux, $d$ is the distance to the target,  $B_\nu (T_{dust})$ is the Planck function at the average disk temperature, and $\kappa_{\nu}$ is the total opacity. Thus, adopting a distance of 310 pc (Section \ref{sec:stellarparameter}) and making standard assumptions concerning the disc temperature (T$_{dust}$ = 20K) and dust opacity ($\kappa_{\nu}$ =  2.3  $\rm cm^{2} g^{-1}$ at 1.33 mm; \citet[and references therein]{Andrews2005}), we estimate disk masses for all detected targets and  report them in Table \ref{Table:observations}.

Similarly, the 3$\sigma$ upper limits of $\sim$0.45 mJy for most of our targets correspond to a dust mass of  M$_{Dust}$ $\sim$ 1.3 $\rm M_{\oplus }$.

\subsection{Target Properties}
\label{sec:stellarparameter}

\begin{figure}
  \centering
     \includegraphics[width=0.47\textwidth]{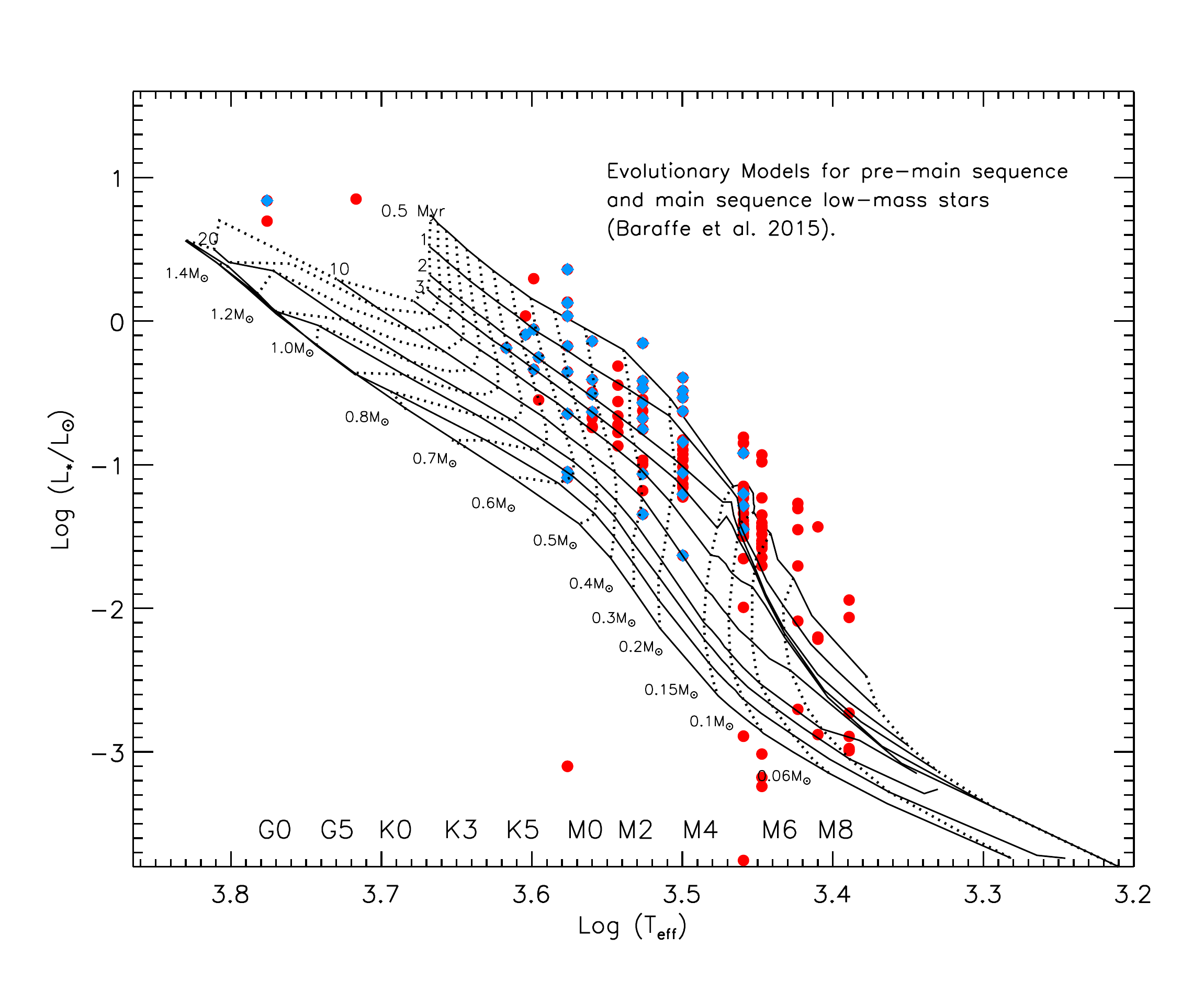} 
   \caption[Theoretical models from \citet{Baraffe2015}]{Inferred stellar parameters for IC 348 members (Table \ref{Table:properties}; Sec. \ref{sec:stellarparameter}) with theoretical models from \citet{Baraffe2015} for low mass young stars overlaid. Solid lines in descending order are 0.5, 1, 2, 3, 5, 10, 20, 50 and 100 Myrs isochrones and dashed lines represent the evolutionary tracks in the range of 0.06 and 1.4 M$_{\odot}$. Blue diamonds represent IC 348 detected members, while red circles correspond to non-detections. }
    \label{fig:baraffe}
\end{figure}

Most of our ALMA targets have fundamental stellar parameters such as extinction, stellar masses, luminosity, effective temperature, etc, reported in previous studies. However, not all values have been obtained in homogeneous manner, and uncertainties might be larger due to systematic differences in methodology or the adopted distance to IC 348. Considering that the most recent data releases of the ${\it Gaia}$ DR2 and Pan-STARRS-1 (PS1) are available, we seek for uniformity in these estimations. We adopt a uniform distance of 310 $\pm$ 20 pc to all targets based on the ${\it Gaia}$ DR2 parallax measurements \citep{Luri2018} of 35 targets in our sample that have DR2 parallax uncertainties smaller than 10$\%$. These objects are indicated with red crosses in Figure~\ref{Fig:area}. To estimate the visual extinction ($\textrm{A}_{\textrm{v}}$), we use the extinction relations $\frac{\textrm{A}_{\lambda_{\textrm{eff}}}}{\textrm{A}_{\textrm{v}}}$ listed in Table \ref{Tab:extinction}, which are calculated using the extinction law presented in \citet{Cardelli1989}. We use PS1 colours r$-$z and z$-$y \citep{Magnier2013}, in order of preference, and adopt the relations $A_{z}=1.4[(r-z)-(r-z)_{0}]$ and $A_{y}=6.3[(z-y)-(z-y)_{0}]$, where $(r-z)_{0}$  and $(z-y)_{0}$ are the expected PS1 colours of a main-sequence star generated from MARCS synthetic fluxes \citep{Gustafsson2008}. In the special case of Class II objects lacking PS1 photometry in the necessary bands, we adopted $A_{v}$ from \citet{Currie2009}.

The stellar luminosities  ($L_{\star}$) of IC~348 members are calculated via the dereddened $J$-band photometry method of \citet{Kenyon1995} and adopting the distance of 310 pc. We derived the stellar properties based on the spectral types taken from \citet{Luhman2003} and \citet{Muench2007} and a conversion from spectral type to the effective temperature (T$_{\rm eff}$) taken from \citet{Pecaut2013} with uncertainties of $\pm$ 1 spectral subclass (Table \ref{Table:properties}). Using T$_{\rm eff}$ and L$_{\star}$, and assuming all targets are single star systems, we estimated the stellar masses ($M_{\star}$) and ages from comparisons with theoretical pre-MS evolutionary tracks. Masses and ages of targets with stellar masses between 0.01 and 1.4 $M_{\odot}$ were derived from models presented in  \citet{Baraffe2015} and stellar masses $>$1.4 $M_{\odot}$ from the PARSEC evolutionary models \citep{Bressan2012}. Age uncertainties are based mainly on the H-R diagram placement and the determination of L$_{\star}$, incorporating the estimated observational photometry, J-band bolometric correction and extinction uncertainties. Nevertheless, the dominant sources of error on the L$_{\star}$ values are the $\sim$ 10$\%$ distance and extinction uncertainties \citep{Cieza2007}. Stellar mass uncertainties are dominated by the $\pm$ 1 spectral subclass and determined by their spectral type. For these T Tauri stars, whose metallicity values are unknown, we adopted solar composition, and we held the surface gravity fixed at the value log $\textrm{g}=$ 4.0, typical for PMS stars. Table \ref{Table:properties} lists the resulting adopted T$_{\rm eff}$, estimated stellar age, $\textrm{A}_{\textrm{v}}$, $L_{\star}$ and estimated stellar mass of these objects. Figure ~\ref{fig:baraffe} shows Baraffe evolutionary models with our IC~348 target selection.  The stars are clustered around the 2 to 3 Myr isochrones, in agreement with previous age estimates for the region.


\begin{table} 
\scriptsize 
\caption{Non-Detected Class II Sources in IC~348}
\label{Table:observations_no}
\begin{tabular}{cccccc}
\toprule
\textbf{Source} & \textbf{$\rm F_{1.3 \, mm}$} & \textbf{rms}& \textbf{Source} & \textbf{$\rm F_{1.3 \, mm}$} & \textbf{rms} \\
\textbf{} & \textbf{[mJy]} & \textbf{[mJy beam$^{-1}$]}& \textbf{} & \textbf{[mJy]} & \textbf{[mJy beam$^{-1}$]} \\
\midrule
1	&	0.19	$\pm$	0.13	&	0.14	&	71	&	0.35	$\pm$	0.13	&	0.13		\\
2	&	0.51	$\pm$	0.13	&	0.13	&	72	&	0.06	$\pm$	0.13	&	0.13		\\
5	&	0.31	$\pm$	0.15	&	0.13	&	73	&	0.40	$\pm$	0.14	&	0.13		\\
6	&	0.52	$\pm$	0.15	&	0.13	&	74	&	0.04	$\pm$	0.13	&	0.14		\\
8	&	0.01	$\pm$	0.14	&	0.14	&	75	&	0.30	$\pm$	0.14	&	0.13		\\
12	&	-0.07	$\pm$	0.14	&	0.14	&	76	&	-0.02	$\pm$	0.14	&	0.14		\\
16	&	-0.05	$\pm$	0.14	&	0.14	&	77	&	-0.19	$\pm$	0.15	&	0.14		\\
18	&	0.17	$\pm$	0.13	&	0.12	&	78	&	0.08	$\pm$	0.14	&	0.13		\\
19	&	0.06	$\pm$	0.15	&	0.13	&	79	&	0.36	$\pm$	0.13	&	0.14		\\
21	&	0.04	$\pm$	0.14	&	0.14	&	80	&	0.27	$\pm$	0.13	&	0.14		\\
22	&	0.12	$\pm$	0.13	&	0.13	&	81	&	0.08	$\pm$	0.13	&	0.14		\\
23	&	-0.09	$\pm$	0.14	&	0.14	&	82	&	-0.11	$\pm$	0.13	&	0.13		\\
24	&	0.13	$\pm$	0.12	&	0.13	&	83	&	0.20	$\pm$	0.14	&	0.13		\\
26	&	0.41	$\pm$	0.14	&	0.14	&	84	&	0.09	$\pm$	0.13	&	0.13		\\
27	&	-0.42	$\pm$	0.13	&	0.14	&	85	&	0.26	$\pm$	0.13	&	0.15		\\
28	&	-0.15	$\pm$	0.13	&	0.14	&	89	&	0.49	$\pm$	0.14	&	0.14		\\
29	&	0.15	$\pm$	0.15	&	0.14	&	93	&	0.26	$\pm$	0.13	&	0.14		\\
30	&	0.26	$\pm$	0.15	&	0.14	&	97	&	0.32	$\pm$	0.14	&	0.12		\\
31	&	0.03	$\pm$	0.13	&	0.14	&	98	&	0.16	$\pm$	0.13	&	0.13		\\
32	&	0.36	$\pm$	0.14	&	0.13	&	99	&	0.25	$\pm$	0.14	&	0.14		\\
33	&	0.16	$\pm$	0.15	&	0.13	&	100	&	0.41	$\pm$	0.15	&	0.13		\\
37	&	0.02	$\pm$	0.12	&	0.14	&	104	&	0.16	$\pm$	0.13	&	0.13		\\
39	&	0.37	$\pm$	0.15	&	0.13	&	105	&	0.21	$\pm$	0.13	&	0.14		\\
40	&	0.31	$\pm$	0.14	&	0.14	&	106	&	0.24	$\pm$	0.12	&	0.13		\\
42	&	0.16	$\pm$	0.13	&	0.14	&	107	&	0.38	$\pm$	0.14	&	0.13		\\
44	&	0.19	$\pm$	0.13	&	0.14	&	108	&	0.36	$\pm$	0.14	&	0.13		\\
45	&	0.35	$\pm$	0.14	&	0.14	&	109	&	-0.31	$\pm$	0.13	&	0.12		\\
46	&	0.24	$\pm$	0.13	&	0.14	&	110	&	0.14	$\pm$	0.14	&	0.14		\\
47	&	-0.13	$\pm$	0.13	&	0.14	&	112	&	-0.38	$\pm$	0.14	&	0.13		\\
49	&	-0.13	$\pm$	0.13	&	0.14	&	113	&	0.21	$\pm$	0.14	&	0.13		\\
50	&	0.20	$\pm$	0.13	&	0.13	&	114	&	0.05	$\pm$	0.13	&	0.12		\\
51	&	0.05	$\pm$	0.13	&	0.13	&	115	&	0.03	$\pm$	0.13	&	0.13		\\
53	&	0.09	$\pm$	0.13	&	0.13	&	117	&	0.25	$\pm$	0.14	&	0.13		\\
54	&	0.50	$\pm$	0.13	&	0.13	&	118	&	0.28	$\pm$	0.14	&	0.15		\\
56	&	0.04	$\pm$	0.14	&	0.14	&	119	&	0.35	$\pm$	0.13	&	0.14		\\
57	&	0.33	$\pm$	0.14	&	0.13	&	120	&	-0.44	$\pm$	0.17	&	0.19		\\
58	&	0.15	$\pm$	0.12	&	0.13	&	122	&	0.02	$\pm$	0.12	&	0.13		\\
59	&	-0.28	$\pm$	0.13	&	0.14	&	123	&	0.21	$\pm$	0.13	&	0.14		\\
60	&	0.20	$\pm$	0.17	&	0.16	&	124	&	0.36	$\pm$	0.13	&	0.14		\\
61	&	-0.18	$\pm$	0.13	&	0.14	&	125	&	0.15	$\pm$	0.14	&	0.13		\\
62	&	0.24	$\pm$	0.12	&	0.13	&	127	&	-0.06	$\pm$	0.15	&	0.14		\\
63	&	0.17	$\pm$	0.14	&	0.12	&	128	&	-0.04	$\pm$	0.13	&	0.14		\\
64	&	0.14	$\pm$	0.13	&	0.14	&	130	&	0.32	$\pm$	0.13	&	0.13		\\
65	&	0.02	$\pm$	0.14	&	0.14	&	131	&	-0.08	$\pm$	0.14	&	0.14		\\
66	&	-0.18	$\pm$	0.13	&	0.14	&	133	&	0.05	$\pm$	0.13	&	0.13		\\
68	&	0.45	$\pm$	0.15	&	0.13	&	134	&	-0.06	$\pm$	0.12	&	0.15		\\
69	&	0.37	$\pm$	0.13	&	0.14	&	135	&	0.07	$\pm$	0.13	&	0.13		\\
70	&	0.15	$\pm$	0.14	&	0.14	&	136	&	0.18	$\pm$	0.14	&	0.14		\\
\bottomrule
\end{tabular}
\end{table}

\begin{table*}
 \centering
 \begin{minipage}{120mm}
  \caption{Continuum Detections of Class II Sources in IC~348}
 \label{Table:observations}
  \begin{tabular}{ccccccc}
 \hline \hline 
 \multicolumn{1}{c}{\textbf{Source}} &
   \multicolumn{1}{c}{\textbf{$\rm F_{1.3 \, mm}$}\footnotemark[1]} &
   \multicolumn{1}{c}{\textbf{rms}} &
   \multicolumn{1}{c}{\textbf{$\Delta \alpha$}} &
   \multicolumn{1}{c}{\textbf{$\Delta \delta$}}  &
    \multicolumn{1}{c}{\textbf{a}}  &
   \multicolumn{1}{c}{\textbf{M$_{Dust}$}} \\[0.5ex] 
    \multicolumn{1}{c}{\textbf{}} &
   \multicolumn{1}{c}{\textbf{[mJy]}} &
   \multicolumn{1}{c}{\textbf{[mJy beam$^{-1}$]}} &
  \multicolumn{1}{c}{\textbf{[\rm $''$]}} &
   \multicolumn{1}{c}{\textbf{[\rm $''$]}} &
   \multicolumn{1}{c}{\textbf{[\rm $''$]}} &
     \multicolumn{1}{c}{\textbf{[M$_{\oplus}$]}} \\[1.5ex]\hline \hline \\[-2ex]

 3	&	2.50	$\pm$	0.24	&	0.14	&	0.19	&	-0.16	&	0	&			6.99	$\pm$	0.30	\\
4	&	1.28	$\pm$	0.15	&	0.13	&	0.06	&	-0.13	&	0	&			3.58	$\pm$	0.28	\\
7	&	5.68	$\pm$	0.45	&	0.13	&	-0.09	&	-0.08	&	0	&			15.88 $\pm$	0.28	\\
9	&	0.70	$\pm$	0.14	&	0.13	&	-0.02	&	-0.10	&	0	&			1.96	$\pm$	0.28	\\
10	&	3.23	$\pm$	0.29	&	0.14	&	0.12	&	-0.06	&	0	&			9.03	$\pm$	0.30	\\
11	&	0.96	$\pm$	0.14	&	0.13	&	0.11	&	-0.12	&	0	&			2.68	$\pm$	0.28	\\
13	&	54.18 $\pm$	4.41	&	0.27	&	0.14	&	0.03	&	0.21  $\pm$ 0.01&	151.82 $\pm$	0.57	\\
14	&	0.78	$\pm$	0.22	&	0.13	&	-0.02	&	-0.14	&	0	&			2.18	$\pm$	0.28	\\
15	&	2.24	$\pm$	0.22	&	0.14	&	-0.13	&	-0.08	&	0	&			6.26	$\pm$	0.30	\\
17	&	1.55	$\pm$	0.17	&	0.14	&	0.07	&	-0.04	&	0	&			4.33	$\pm$	0.30	\\
20	&	0.78	$\pm$	0.15	&	0.13	&	-0.34	&	-0.40	&	0	&			2.18	$\pm$	0.28	\\
25	&	1.22	$\pm$	0.17	&	0.13	&	0.01	&	-0.20	&	0	&			3.41	$\pm$	0.28	\\
34	&	4.09	$\pm$	0.36	&	0.13	&	-0.14	&	-0.28	&	0.23 $\pm$ 0.04&	11.43 $\pm$	0.28	\\
35	&	6.34	$\pm$	0.54	&	0.13	&	-0.01	&	-0.15	&	0.33 $\pm$ 0.02&	17.72 $\pm$	0.28	\\
36	&	1.74	$\pm$	0.20	&	0.14	&	-0.09	&	-0.13	&	0	&			4.86	$\pm$	0.30	\\
38	&	1.42	$\pm$	0.17	&	0.13	&	-0.23	&	-0.25	&	0	&			3.97	$\pm$	0.28	\\
41	&	0.61	$\pm$	0.13	&	0.13	&	0.17	&	0.18	&	0	&			1.70	$\pm$	0.28	\\
43	&	0.59	$\pm$	0.14	&	0.14	&	-0.06	&	-0.06	&	0	&			1.65	$\pm$	0.30	\\
48	&	2.08	$\pm$	0.2	&	0.13	&	0.10	&	-0.03	&	0	&			5.81	$\pm$	0.28	\\
52	&	0.67	$\pm$	0.14	&	0.14	&	0.10	&	-0.04	&	0	&			1.87	$\pm$	0.30	\\
55	&	0.80	$\pm$	0.15	&	0.14	&	0.10	&	-0.17	&	0	&			2.24	$\pm$	0.30	\\
67	&	1.91	$\pm$	0.21	&	0.13	&	0.13	&	-0.04	&	0	&			5.34	$\pm$	0.28	\\
86	&	1.27	$\pm$	0.17	&	0.13	&	-0.10	&	-0.31	&	0	&			3.55	$\pm$	0.28	\\
87	&	0.70	$\pm$	0.14	&	0.13	&	-0.02	&	-0.13	&	0	&			1.96	$\pm$	0.28	\\
88	&	2.99	$\pm$	0.28	&	0.13	&	-0.02	&	-0.03	&	0	&			8.36	$\pm$	0.28	\\
90	&	1.69	$\pm$	0.20	&	0.15	&	0.11	&	-0.08	&	0.37 $\pm$ 0.08&	4.72	$\pm$	0.32	\\
91	&	6.73	$\pm$	0.58	&	0.15	&	0.08	&	-0.17	&	0.38 $\pm$  0.02&	18.81 $\pm$	0.32	\\
92	&	12.01	$\pm$	1.12	&	0.16	&	0.09	&	-0.09	&	0.47 $\pm$ 0.01&33.57 $\pm$	0.34	\\
94	&	6.00	$\pm$	0.50	&	0.14	&	0.05	&	-0.27	&	0.31 $\pm$ 0.02&	16.77 $\pm$	0.30	\\
95	&	1.88	$\pm$	0.22	&	0.14	&	0.09	&	-0.26	&	0	&			5.25	$\pm$	0.30	\\
96	&	1.76	$\pm$	0.19	&	0.14	&	0.10	&	-0.10	&	0	&			4.92	$\pm$	0.30	\\
101	&	2.50	$\pm$	0.24	&	0.14	&	-0.01	&	-0.06	&	0	&			6.99	$\pm$	0.30	\\
102	&	0.73	$\pm$	0.14	&	0.14	&	0.20	&	0.01	&	0	&			2.04	$\pm$	0.30	\\
103	&	7.22	$\pm$	0.58	&	0.14	&	0.02	&	-0.06	&	0.30 $\pm$ 0.02&    20.18	$\pm$	0.30	\\
111	&	1.06	$\pm$	0.16	&	0.13	&	-0.02	&	-0.01	&	0	&			2.96	$\pm$	0.28	\\
116	&	1.43	$\pm$	0.16	&	0.13	&	0.04	&	-0.08	&	0	&			4.03	$\pm$	0.28	\\
121	&	4.83	$\pm$	0.41	&	0.14	&	0.11	&	0	&	0.22 $\pm$ 0.04&	13.50 $\pm$	0.30	\\
126	&	6.54	$\pm$	0.52	&	0.13	&	0	&	-0.04	&	0	&			18.28 $\pm$	0.28	\\
129	&	0.62	$\pm$	0.15	&	0.12	&	0.18	&	-0.29	&	0.39 $\pm$  0.17&	1.73	$\pm$	0.25	\\
132	&	1.05	$\pm$	0.15	&	0.13	&	0.17	&	-0.17	&	0	&                       2.93 $\pm$ 0.28\\[0.1ex]\hline
       
\end{tabular}
\begin{tablenotes}
\item[*] $^{*}$The elliptical Gaussian model applied to the resolved sources generates five free parameters. Here, we report: integrated flux density ($\rm F_{1.33 mm}$), FWHM along the major axis (a), right ascension offset from the phase center (${\Delta \alpha}$), and declination offset from the phase center (${\Delta \delta}$).
 \end{tablenotes}
\end{minipage}
\end{table*}

\begin{table*}
\centering
\begin{tabular}{cllc}
\hline
Filter ID    & $\textrm{A}_{\lambda_{\textrm{eff}}}$ & $\frac{\textrm{A}_{\lambda_{\textrm{eff}}}}{\textrm{A}_{\textrm{v}}}$ & Zero Point \\
& [\AA] &  & [Jy] \\
\hline
g     & 4775.6    & 1.19 & 3631     \\
 r        & 6129.5       & 0.89 & 3631       \\
i       & 7484.6     & 0.67 & 3631     \\
z       & 8657.8	    & 0.51 & 3631     \\
y & 9603.1   & 0.44   & 3631    \\
\hline
\end{tabular}
\caption{Extinction relations calculated by following the \citet{Cardelli1989} extinct law with R$_{v}$ = 3.1. }
\label{Tab:extinction}
\end{table*}

\onecolumn

\clearpage
\renewcommand{\thefootnote}{\fnsymbol{footnote}}
\begin{spacing}{1.5}
\begin{longtable}{cccccccccccl}

\caption[IC 348 Stellar Properties]{Stellar Properties for Class II Sources in IC~348.}
\label{Table:properties} \\
\hline \hline \\[-0.9ex]
   \multicolumn{1}{c}{\textbf{Source ID}} &
   \multicolumn{1}{c}{\textbf{Log T$_{\rm eff}$ }} &
   \multicolumn{1}{c}{\textbf{$\textrm{A}_{\textrm{v}}$}} &
   \multicolumn{1}{c}{\textbf{Log L$_{\star}$}} &
    \multicolumn{1}{c}{\textbf{M$_{\star}$}}&
    \multicolumn{1}{c}{\textbf{Source ID}} &
   \multicolumn{1}{c}{\textbf{Log T$_{\rm eff}$ }} &
   \multicolumn{1}{c}{\textbf{$\textrm{A}_{\textrm{v}}$}} &
   \multicolumn{1}{c}{\textbf{Log L$_{\star}$}} &
    \multicolumn{1}{c}{\textbf{M$_{\star}$}} \\[0.6ex] 
    \multicolumn{1}{c}{\textbf{}} &
   \multicolumn{1}{c}{\textbf{[K]}} &
   \multicolumn{1}{c}{\textbf{[mag]}} &
  \multicolumn{1}{c}{\textbf{[L$_{\odot}$]}} &
   \multicolumn{1}{c}{\textbf{[M$_{\odot}$]}} &
   \multicolumn{1}{c}{\textbf{}} &
   \multicolumn{1}{c}{\textbf{[K]}} &
   \multicolumn{1}{c}{\textbf{[mag]}} &
  \multicolumn{1}{c}{\textbf{[L$_{\odot}$]}} &
   \multicolumn{1}{c}{\textbf{[M$_{\odot}$]}} &
     \\[1.5ex]\hline \hline \\[-1ex]
\endfirsthead

\multicolumn{6}{c}{{\tablename} \thetable{} -- Continued}\\[1.0ex]
  \hline \hline \\[-0.9ex]
   \multicolumn{1}{c}{\textbf{Source ID}} &
   \multicolumn{1}{c}{\textbf{Log T$_{\rm eff}$ }} &
   \multicolumn{1}{c}{\textbf{$\textrm{A}_{\textrm{v}}$}} &
   \multicolumn{1}{c}{\textbf{Log L$_{\star}$}} &
    \multicolumn{1}{c}{\textbf{M$_{\star}$}}&
    \multicolumn{1}{c}{\textbf{Source ID}} &
   \multicolumn{1}{c}{\textbf{Log T$_{\rm eff}$ }} &
   \multicolumn{1}{c}{\textbf{$\textrm{A}_{\textrm{v}}$}} &
   \multicolumn{1}{c}{\textbf{Log L$_{\star}$}} &
    \multicolumn{1}{c}{\textbf{M$_{\star}$}} \\[0.6ex] 
    \multicolumn{1}{c}{\textbf{}} &
   \multicolumn{1}{c}{\textbf{[K]}} &
   \multicolumn{1}{c}{\textbf{[mag]}} &
  \multicolumn{1}{c}{\textbf{[L$_{\odot}$]}} &
   \multicolumn{1}{c}{\textbf{[M$_{\odot}$]}} &
   \multicolumn{1}{c}{\textbf{}} &
   \multicolumn{1}{c}{\textbf{[K]}} &
   \multicolumn{1}{c}{\textbf{[mag]}} &
  \multicolumn{1}{c}{\textbf{[L$_{\odot}$]}} &
   \multicolumn{1}{c}{\textbf{[M$_{\odot}$]}} &
     \\[1.5ex]\hline \hline \\[-1ex]

   \endhead
\endfoot

\hline\hline\\[-0.5ex] 
\endlastfoot


1	&	3.95$_{-0.01}^{+0.02}$	&	2.87	$\pm$	0.08	&	2.00	$	_{-0.19}^{+0.16}	$	&	3.17$_{-0.00}^{+0.00}$	&	69	&	3.45$_{-0.02}^{+0.01}$	&	2.85	$\pm$	0.16	&	-1.70	$	_{-0.38}^{+0.24}		$	&	0.04$_{-0.01}^{+0.01}$	\\
2	&	3.72$_{-0.01}^{+0.01}$	&	4.60	$\pm$	0.03	&	0.85	$	_{-0.21}^{+0.17}	$	&	1.80$_{-0.00}^{+0.00}$	&	70	&	3.45$_{-0.02}^{+0.01}$	&	4.18	$\pm$	0.11	&	-1.55	$	_{-0.38}^{+0.24}		$	&	0.05$_{-0.01}^{+0.01}$	\\
3	&	3.58$_{-0.02}^{+0.01}$	&	13.93	$\pm$	1.89	&	0.36	$	_{-0.26}^{+0.20}	$	&	0.47$_{-0.04}^{+0.03}$	&	71	&	3.53$_{-0.03}^{+0.02}$	&	9.34	$\pm$	0.24	&	-1.18	$	_{-0.33}^{+0.24}		$	&	0.30$_{-0.07}^{+0.05}$	\\
4	&	3.58$_{-0.02}^{+0.01}$	&	2.22	$\pm$	0.03	&	0.04	$	_{-0.30}^{+0.23}	$	&	0.48$_{-0.04}^{+0.03}$	&	72	&	3.41$_{-0.02}^{+0.01}$	&	4.38	$\pm$	0.49	&	-2.20	$	_{-0.93}^{+0.27}		$	&	0.02$_{-0.01}^{+0.01}$	\\
5	&	3.95$_{-0.01}^{+0.02}$	&	1.30	$\pm$	0.01	&	1.08	$	_{-0.19}^{+0.16}	$	&	1.89$_{-0.00}^{+0.00}$	&	73	&	3.45$_{-0.02}^{+0.01}$	&	1.70	$\pm$	0.08	&	-1.48	$	_{-0.38}^{+0.24}		$	&	0.05$_{-0.01}^{+0.01}$	\\
6	&	3.60$_{-0.00}^{+0.01}$	&	8.99	$\pm$	0.36	&	0.30	$	_{-0.20}^{+0.18}	$	&	0.60$_{-0.01}^{+0.02}$	&	74	&	3.41$_{-0.02}^{+0.01}$	&	4.11	$\pm$	0.89	&	-2.22	$	_{-0.93}^{+0.27}		$	&	0.02$_{-0.01}^{+0.01}$	\\
7	&	3.60$_{-0.00}^{+0.01}$	&	3.60	$\pm$	0.06	&	-0.06	$	_{-0.21}^{+0.19}	$	&	0.61$_{-0.01}^{+0.02}$	&	75	&	3.46$_{-0.01}^{+0.04}$	&	5.40	$\pm$	0.10	&	-2.89	$	_{-0.19}^{+0.20}		$	&	0.07$_{-0.01}^{+0.04}$	\\
8	&	3.60$_{-0.01}^{+0.01}$	&	2.52	$\pm$	0.06	&	0.04	$	_{-0.22}^{+0.21}	$	&	0.65$_{-0.02}^{+0.05}$	&	76	&	3.39$_{-0.00}^{+0.00}$	&	1.60	$\pm$	0.10	&	-2.98	$	_{-0.19}^{+0.16}		$	&	0.02$_{-0.01}^{+0.01}$	\\
9	&	3.60$_{-0.01}^{+0.01}$	&	2.32	$\pm$	0.12	&	-0.09	$	_{-0.22}^{+0.21}	$	&	0.69$_{-0.02}^{+0.05}$	&	77	&	3.39$_{-0.00}^{+0.02}$	&	4.39	$\pm$	0.53	&	-2.06	$	_{-1.03}^{+0.26}		$	&	0.01$_{-0.01}^{+0.01}$	\\
10	&	3.60$_{-0.01}^{+0.00}$	&	2.67	$\pm$	0.09	&	-0.25	$	_{-0.24}^{+0.17}	$	&	0.61$_{-0.02}^{+0.01}$	&	78	&	3.41$_{-0.02}^{+0.01}$	&	3.87	$\pm$	0.34	&	-1.43	$	_{-0.94}^{+0.27}		$	&	0.03$_{-0.10}^{+0.01}$	\\
11	&	3.60$_{-0.00}^{+0.01}$	&	3.94	$\pm$	0.12	&	-0.34	$	_{-0.21}^{+0.19}	$	&	0.69$_{-0.01}^{+0.02}$	&	79	&	3.45$_{-0.00}^{+0.00}$	&	1.60	$\pm$	0.10	&	-3.01	$	_{-0.19}^{+0.16}		$	&	0.06$_{-0.00}^{+0.00}$	\\
12	&	3.58$_{-0.02}^{+0.01}$	&	9.44	$\pm$	0.07	&	0.13	$	_{-0.26}^{+0.20}	$	&	0.45$_{-0.04}^{+0.03}$	&	80	&	3.39$_{-0.00}^{+0.00}$	&	1.50	$\pm$	0.10	&	-2.73	$	_{-0.19}^{+0.16}		$	&	0.06$_{-0.00}^{+0.00}$	\\
13	&	3.58$_{-0.02}^{+0.01}$	&	10.78	$\pm$	1.03	&	-0.65	$	_{-0.27}^{+0.25}	$	&	0.59$_{-0.04}^{+0.03}$	&	81	&	3.46$_{-0.00}^{+0.00}$	&	1.80	$\pm$	0.10	&	-2.89	$	_{-0.19}^{+0.16}		$	&	0.07$_{-0.00}^{+0.00}$	\\
14	&	3.58$_{-0.02}^{+0.01}$	&	12.46	$\pm$	0.23	&	0.13	$	_{-0.26}^{+0.20}	$	&	0.45$_{-0.04}^{+0.03}$	&	82	&	3.42$_{-0.00}^{+0.00}$	&	0.00	$\pm$	0.0	&	-2.70	$	_{-0.19}^{+0.16}		$	&	0.04$_{-0.00}^{+0.00}$	\\
15	&	3.56$_{-0.02}^{+0.02}$	&	1.74	$\pm$	0.06	&	-0.41	$	_{-0.31}^{+0.26}	$	&	0.41$_{-0.04}^{+0.05}$	&	83	&	3.45$_{-0.00}^{+0.00}$	&	1.10	$\pm$	0.10	&	-3.24	$	_{-0.19}^{+0.16}		$	&	0.06$_{-0.00}^{+0.00}$	\\
16	&	3.60$_{-0.01}^{+0.00}$	&	2.26	$\pm$	0.18	&	-0.55	$	_{-0.24}^{+0.17}	$	&	0.70$_{-0.02}^{+0.01}$	&	84	&	3.39$_{-0.00}^{+0.00}$	&	1.30	$\pm$	0.10	&	-2.89	$	_{-0.19}^{+0.16}		$	&	0.05$_{-0.00}^{+0.00}$	\\
17	&	3.56$_{-0.02}^{+0.02}$	&	1.60	$\pm$	0.11	&	-0.50	$	_{-0.31}^{+0.26}	$	&	0.41$_{-0.04}^{+0.05}$	&	85	&	3.78$_{-0.01}^{+0.01}$	&	5.94	$\pm$	0.14	&	0.70	$	_{-0.21}^{+0.17}		$	&	1.51$_{-0.00}^{+0.00}$	\\
18	&	3.53$_{-0.03}^{+0.02}$	&	1.06	$\pm$	0.06	&	-0.55	$	_{-0.34}^{+0.26}	$	&	0.27$_{-0.04}^{+0.03}$	&	86	&	3.50$_{-0.04}^{+0.03}$	&	1.93	$\pm$	0.04	&	-0.39	$	_{-0.38}^{+0.27}		$	&	0.18$_{-0.04}^{+0.04}$	\\
19	&	3.56$_{-0.02}^{+0.02}$	&	0.0	$\pm$	0.0	&	-0.74	$	_{-0.22}^{+0.18}	$	&	0.48$_{-0.05}^{+0.05}$	&	87	&	3.58$_{-0.02}^{+0.01}$	&	7.48	$\pm$	0.36	&	-0.17	$	_{-0.26}^{+0.20}		$	&	0.49$_{-0.04}^{+0.03}$	\\
20	&	3.53$_{-0.03}^{+0.02}$	&	1.69	$\pm$	0.06	&	-0.57	$	_{-0.35}^{+0.26}	$	&	0.27$_{-0.05}^{+0.03}$	&	88	&	3.58$_{-0.02}^{+0.01}$	&	2.55	$\pm$	0.04	&	-0.35	$	_{-0.30}^{+0.23}		$	&	0.50$_{-0.05}^{+0.04}$	\\
21	&	3.56$_{-0.02}^{+0.02}$	&	1.27	$\pm$	0.20	&	-0.68	$	_{-0.31}^{+0.26}	$	&	0.44$_{-0.04}^{+0.05}$	&	89	&	3.54$_{-0.02}^{+0.02}$	&	2.53	$\pm$	0.04	&	-0.56	$	_{-0.27}^{+0.24}		$	&	0.35$_{-0.03}^{+0.04}$	\\
22	&	3.54$_{-0.02}^{+0.02}$	&	1.55	$\pm$	0.11	&	-0.66	$	_{-0.30}^{+0.28}	$	&	0.35$_{-0.03}^{+0.04}$	&	90	&	3.50$_{-0.04}^{+0.03}$	&	1.61	$\pm$	0.10	&	-0.53	$	_{-0.38}^{+0.27}		$	&	0.18$_{-0.04}^{+0.04}$	\\
23	&	3.54$_{-0.02}^{+0.02}$	&	5.37	$\pm$	0.6	&	-0.45	$	_{-0.27}^{+0.24}	$	&	0.36$_{-0.03}^{+0.04}$	&	91	&	3.56$_{-0.02}^{+0.02}$	&	1.18	$\pm$	0.02	&	-0.14	$	_{-0.31}^{+0.26}		$	&	0.39$_{-0.03}^{+0.04}$	\\
24	&	3.50$_{-0.04}^{+0.03}$	&	0.43	$\pm$	0.13	&	-0.96	$	_{-0.28}^{+0.31}	$	&	0.17$_{-0.06}^{+0.04}$	&	92	&	3.78$_{-0.01}^{+0.01}$	&	10.90	$\pm$	0.08	&	0.84	$	_{-0.22}^{+0.18}		$	&	1.56$_{-0.00}^{+0.00}$	\\
25	&	3.56$_{-0.02}^{+0.02}$	&	1.93	$\pm$	0.07	&	-0.63	$	_{-0.31}^{+0.26}	$	&	0.45$_{-0.05}^{+0.05}$	&	93	&	3.42$_{-0.01}^{+0.02}$	&	2.87	$\pm$	0.14	&	-1.45	$	_{-0.31}^{+0.34}		$	&	0.04$_{-0.00}^{+0.01}$	\\
26	&	3.54$_{-0.02}^{+0.02}$	&	1.18	$\pm$	0.08	&	-0.87	$	_{-0.30}^{+0.28}	$	&	0.39$_{-0.04}^{+0.05}$	&	94	&	3.56$_{-0.02}^{+0.02}$	&	1.22	$\pm$	0.05	&	-0.51	$	_{-0.28}^{+0.23}		$	&	0.41$_{-0.04}^{+0.05}$	\\
27	&	3.54$_{-0.02}^{+0.02}$	&	6.90	$\pm$	0.17	&	-0.31	$	_{-0.27}^{+0.24}	$	&	0.33$_{-0.03}^{+0.03}$	&	95	&	3.53$_{-0.03}^{+0.02}$	&	1.96	$\pm$	0.07	&	-0.47	$	_{-0.31}^{+0.23}		$	&	0.26$_{-0.05}^{+0.03}$	\\
28	&	3.54$_{-0.02}^{+0.02}$	&	1.87	$\pm$	0.02	&	-0.72	$	_{-0.30}^{+0.28}	$	&	0.36$_{-0.04}^{+0.04}$	&	96	&	3.53$_{-0.03}^{+0.02}$	&	1.76	$\pm$	0.12	&	-1.06	$	_{-0.31}^{+0.23}		$	&	0.30$_{-0.06}^{+0.04}$	\\
29	&	3.54$_{-0.02}^{+0.02}$	&	1.24	$\pm$	0.02	&	-0.78	$	_{-0.30}^{+0.28}	$	&	0.37$_{-0.04}^{+0.04}$	&	97	&	3.53$_{-0.03}^{+0.02}$	&	0.66	$\pm$	0.32	&	-0.97	$	_{-0.29}^{+0.26}		$	&	0.29$_{-0.06}^{+0.04}$	\\
30	&	3.50$_{-0.04}^{+0.03}$	&	1.10	$\pm$	0.02	&	-0.83	$	_{-0.36}^{+0.31}	$	&	0.18$_{-0.06}^{+0.04}$	&	98	&	3.50$_{-0.04}^{+0.03}$	&	0.00	$\pm$	0	&	-1.09	$	_{-0.23}^{+0.28}		$	&	0.17$_{-0.06}^{+0.05}$	\\
31	&	3.46$_{-0.01}^{+0.04}$	&	2.42	$\pm$	0.15	&	-0.81	$	_{-0.27}^{+0.40}	$	&	0.06$_{-0.02}^{+0.04}$	&	99	&	3.53$_{-0.03}^{+0.02}$	&	4.01	$\pm$	0.10	&	-0.75	$	_{-0.33}^{+0.24}		$	&	0.27$_{-0.05}^{+0.03}$	\\
32	&	3.53$_{-0.03}^{+0.02}$	&	4.69	$\pm$	0.7	&	-1.00	$	_{-0.33}^{+0.24}	$	&	0.30$_{-0.06}^{+0.04}$	&	100	&	3.42$_{-0.01}^{+0.02}$	&	7.88	$\pm$	0.31	&	-1.31	$	_{-0.31}^{+0.34}		$	&	0.05$_{-0.00}^{+0.01}$	\\
33	&	3.46$_{-0.01}^{+0.04}$	&	2.73	$\pm$	0.11	&	-0.85	$	_{-0.27}^{+0.40}	$	&	0.061$_{-0.02}^{+0.04}$	&	101	&	3.53$_{-0.03}^{+0.02}$	&	11.29	$\pm$	0.74	&	-0.75	$	_{-0.33}^{+0.24}		$	&	0.28$_{-0.05}^{+0.04}$	\\
34	&	3.46$_{-0.01}^{+0.04}$	&	2.70	$\pm$	0.08	&	-0.92	$	_{-0.27}^{+0.40}	$	&	0.06$_{-0.02}^{+0.04}$	&	102	&	3.53$_{-0.03}^{+0.02}$	&	4.59	$\pm$	0.82	&	-1.34	$	_{-0.33}^{+0.24}		$	&	0.29$_{-0.06}^{+0.06}$	\\
35	&	3.50$_{-0.04}^{+0.03}$	&	0.41	$\pm$	0.01	&	-1.06	$	_{-0.28}^{+0.31}	$	&	0.18$_{-0.06}^{+0.05}$	&	103	&	3.53$_{-0.03}^{+0.02}$	&	15.43	$\pm$	1.27	&	-0.42	$	_{-0.34}^{+0.23}		$	&	0.27$_{-0.05}^{+0.03}$	\\
36	&	3.53$_{-0.03}^{+0.02}$	&	11.2	$\pm$	1.05	&	-0.15	$	_{-0.33}^{+0.24}	$	&	0.27$_{-0.04}^{+0.03}$	&	104	&	3.46$_{-0.01}^{+0.04}$	&	3.28	$\pm$	0.11	&	-1.23	$	_{-0.27}^{+0.40}		$	&	0.06$_{-0.01}^{+0.06}$	\\
37	&	3.46$_{-0.01}^{+0.04}$	&	3.43	$\pm$	0.10	&	-0.85	$	_{-0.27}^{+0.40}	$	&	0.061$_{-0.02}^{+0.04}$	&	105	&	3.46$_{-0.01}^{+0.04}$	&	1.93	$\pm$	0.07	&	-1.45	$	_{-0.27}^{+0.40}		$	&	0.08$_{-0.01}^{+0.06}$	\\
38	&	3.50$_{-0.04}^{+0.03}$	&	7.78	$\pm$	0.17	&	-0.48	$	_{-0.44}^{+0.29}	$	&	0.18$_{-0.04}^{+0.04}$	&	106	&	3.45$_{-0.02}^{+0.01}$	&	2.59	$\pm$	0.05	&	-1.41	$	_{-0.38}^{+0.24}		$	&	0.05$_{-0.01}^{+0.01}$	\\
39	&	3.50$_{-0.04}^{+0.03}$	&	3.47	$\pm$	0.12	&	-0.90	$	_{-0.44}^{+0.29}	$	&	0.19$_{-0.06}^{+0.04}$	&	107	&	3.46$_{-0.01}^{+0.04}$	&	5.61	$\pm$	0.20	&	-1.15	$	_{-0.27}^{+0.40}		$	&	0.08$_{-0.01}^{+0.05}$	\\
40	&	3.45$_{-0.02}^{+0.01}$	&	2.62	$\pm$	0.10	&	-0.93	$	_{-0.38}^{+0.24}	$	&	0.04$_{-0.01}^{+0.02}$	&	108	&	3.46$_{-0.01}^{+0.04}$	&	5.78	$\pm$	0.15	&	-1.30	$	_{-0.27}^{+0.40}		$	&	0.08$_{-0.01}^{+0.05}$	\\
41	&	3.50$_{-0.04}^{+0.03}$	&	10.26	$\pm$	0.15	&	-0.63	$	_{-0.44}^{+0.29}	$	&	0.18$_{-0.05}^{+0.04}$	&	109	&	3.45$_{-0.02}^{+0.01}$	&	3.72	$\pm$	0.15	&	-1.53	$	_{-0.38}^{+0.24}		$	&	0.05$_{-0.01}^{+0.01}$	\\
42	&	3.50$_{-0.04}^{+0.03}$	&	2.71	$\pm$	0.13	&	-0.93	$	_{-0.44}^{+0.29}	$	&	0.18$_{-0.06}^{+0.04}$	&	110	&	3.45$_{-0.02}^{+0.01}$	&	5.57	$\pm$	0.28	&	-1.58	$	_{-0.38}^{+0.24}		$	&	0.05$_{-0.01}^{+0.01}$	\\
43	&	3.53$_{-0.03}^{+0.02}$	&	12.2	$\pm$	1.42	&	-0.68	$	_{-0.33}^{+0.24}	$	&	0.27$_{-0.04}^{+0.03}$	&	111	&	3.50$_{-0.04}^{+0.03}$	&	8.21	$\pm$	0.50	&	-1.21	$	_{-0.44}^{+0.29}		$	&	0.18$_{-0.06}^{+0.06}$	\\
44	&	3.56$_{-0.02}^{+0.02}$	&	12.52	$\pm$	1.83	&	-0.73	$	_{-0.26}^{+0.22}	$	&	0.48$_{-0.05}^{+0.05}$	&	112	&	3.45$_{-0.02}^{+0.01}$	&	0.00	$\pm$	0.00	&	-3.18	$	_{-0.19}^{+0.16}		$	&	0.06$_{-0.01}^{+0.00}$	\\
45	&	3.45$_{-0.02}^{+0.01}$	&	3.79	$\pm$	0.06	&	-0.98	$	_{-0.38}^{+0.24}	$	&	0.05$_{-0.01}^{+0.02}$	&	113	&	3.45$_{-0.02}^{+0.01}$	&	14.30	$\pm$	0.01	&	-1.65	$	_{-0.19}^{+0.16}		$	&	0.05$_{-0.01}^{+0.01}$	\\
46	&	3.46$_{-0.01}^{+0.04}$	&	0.79	$\pm$	0.07	&	-1.20	$	_{-0.27}^{+0.40}	$	&	0.06$_{-0.01}^{+0.05}$	&	114	&	3.39$_{-0.00}^{+0.00}$	&	0.00	$\pm$	0.00	&	-2.99	$	_{-0.19}^{+0.16}		$	&	0.03$_{-0.00}^{+0.00}$	\\
47	&	3.50$_{-0.04}^{+0.03}$	&	4.19	$\pm$	0.25	&	-1.02	$	_{-0.44}^{+0.29}	$	&	0.18$_{-0.06}^{+0.05}$	&	115	&	3.53$_{-0.03}^{+0.02}$	&	6.86	$\pm$	0.21	&	-0.62	$	_{-0.33}^{+0.24}		$	&	0.28$_{-0.04}^{+0.03}$	\\
48	&	3.58$_{-0.02}^{+0.01}$	&	6.89	$\pm$	0.75	&	-1.09	$	_{-0.26}^{+0.20}	$	&	0.60$_{-0.03}^{+0.02}$	&	116	&	3.46$_{-0.01}^{+0.04}$	&	2.43	$\pm$	0.12	&	-1.29	$	_{-0.27}^{+0.40}		$	&	0.08$_{-0.01}^{+0.06}$	\\
49	&	3.46$_{-0.01}^{+0.04}$	&	1.20	$\pm$	0.10	&	-1.19	$	_{-0.27}^{+0.40}	$	&	0.06$_{-0.01}^{+0.05}$	&	117	&	3.42$_{-0.01}^{+0.02}$	&	1.42	$\pm$	0.09	&	-1.70	$	_{-0.31}^{+0.34}		$	&	0.03$_{-0.00}^{+0.02}$	\\
50	&	3.46$_{-0.01}^{+0.04}$	&	1.36	$\pm$	0.11	&	-1.18	$	_{-0.27}^{+0.40}	$	&	0.08$_{-0.01}^{+0.05}$	&	118	&	3.46$_{-0.01}^{+0.04}$	&	4.48	$\pm$	0.15	&	-1.42	$	_{-0.27}^{+0.40}		$	&	0.07$_{-0.01}^{+0.06}$	\\
51	&	3.50$_{-0.04}^{+0.03}$	&	4.50	$\pm$	0.11	&	-0.90	$	_{-0.44}^{+0.29}	$	&	0.18$_{-0.06}^{+0.04}$	&	119	&	3.46$_{-0.01}^{+0.04}$	&	2.45	$\pm$	0.13	&	-1.15	$	_{-0.27}^{+0.40}		$	&	0.09$_{-0.01}^{+0.06}$	\\
52	&	3.46$_{-0.01}^{+0.04}$	&	2.69	$\pm$	0.13	&	-1.20	$	_{-0.27}^{+0.40}	$	&	0.08$_{-0.01}^{+0.05}$	&	120	&	3.39$_{-0.00}^{+0.02}$	&	0.00	$\pm$	0.00	&	-1.94	$	_{-0.19}^{+0.16}		$	&	0.02$_{-0.00}^{+0.00}$	\\
53	&	3.45$_{-0.02}^{+0.01}$	&	1.57	$\pm$	0.09	&	-1.23	$	_{-0.37}^{+0.24}	$	&	0.05$_{-0.02}^{+0.02}$	&	121	&	3.50$_{-0.04}^{+0.03}$	&	4.03	$\pm$	0.27	&	-0.84	$	_{-0.44}^{+0.29}		$	&	0.19$_{-0.06}^{+0.04}$	\\
54	&	3.50$_{-0.04}^{+0.03}$	&	5.52	$\pm$	0.10	&	-0.97	$	_{-0.44}^{+0.29}	$	&	0.17$_{-0.06}^{+0.04}$	&	122	&	3.50$_{-0.04}^{+0.03}$	&	3.54	$\pm$	0.05	&	-1.14	$	_{-0.44}^{+0.29}		$	&	0.18$_{-0.06}^{+0.05}$	\\
55	&	3.58$_{-0.02}^{+0.01}$	&	9.65	$\pm$	0.40	&	-1.05	$	_{-0.26}^{+0.20}	$	&	0.61$_{-0.03}^{+0.02}$	&	123	&	3.56$_{-0.02}^{+0.02}$	&	0.28	$\pm$	0.14	&	-0.50	$	_{-0.25}^{+0.26}		$	&	0.43$_{-0.04}^{+0.05}$	\\
56	&	3.42$_{-0.01}^{+0.02}$	&	2.45	$\pm$	0.12	&	-1.27	$	_{-0.30}^{+0.34}	$	&	0.03$_{-0.01}^{+0.01}$	&	124	&	3.46$_{-0.01}^{+0.04}$	&	0.00	$\pm$	0.00	&	-3.76	$	_{-0.19}^{+0.20}		$	&	0.04$_{-0.00}^{+0.04}$	\\
57	&	3.45$_{-0.02}^{+0.01}$	&	2.26	$\pm$	0.08	&	-1.35	$	_{-0.38}^{+0.24}	$	&	0.05$_{-0.02}^{+0.02}$	&	125	&	3.46$_{-0.01}^{+0.04}$	&	0.00	$\pm$	0.00	&	-1.99	$	_{-0.19}^{+0.20}		$	&	0.07$_{-0.01}^{+0.05}$	\\
58	&	3.50$_{-0.04}^{+0.03}$	&	10.56	$\pm$	0.36	&	-0.86	$	_{-0.44}^{+0.29}	$	&	0.18$_{-0.06}^{+0.04}$	&	126	&	3.62$_{-0.01}^{+0.02}$	&	3.42	$\pm$	0.25	&	-0.19	$	_{-0.20}^{+0.18}		$	&	0.80$_{-0.05}^{+0.08}$	\\
59	&	3.46$_{-0.01}^{+0.04}$	&	1.26	$\pm$	0.15	&	-1.44	$	_{-0.27}^{+0.40}	$	&	0.07$_{-0.01}^{+0.06}$	&	127	&	3.46$_{-0.01}^{+0.04}$	&	2.70	$\pm$	0.08	&	-1.39	$	_{-0.27}^{+0.40}		$	&	0.08$_{-0.01}^{+0.06}$	\\
60	&	3.46$_{-0.01}^{+0.04}$	&	2.40	$\pm$	0.09	&	-1.34	$	_{-0.27}^{+0.40}	$	&	0.07$_{-0.01}^{+0.06}$	&	128	&	3.50$_{-0.04}^{+0.03}$	&	2.10	$\pm$	0.09	&	-1.16	$	_{-0.44}^{+0.29}		$	&	0.18$_{-0.06}^{+0.05}$	\\
61	&	3.45$_{-0.02}^{+0.01}$	&	2.90	$\pm$	0.13	&	-1.43	$	_{-0.38}^{+0.24}	$	&	0.05$_{-0.01}^{+0.01}$	&	129	&	3.50$_{-0.04}^{+0.03}$	&	6.45	$\pm$	0.54	&	-1.63	$	_{-0.44}^{+0.29}		$	&	0.15$_{-0.05}^{+0.07}$	\\
62	&	3.45$_{-0.02}^{+0.01}$	&	3.17	$\pm$	0.19	&	-1.44	$	_{-0.38}^{+0.24}	$	&	0.05$_{-0.01}^{+0.01}$	&	130	&	3.50$_{-0.04}^{+0.03}$	&	6.72	$\pm$	0.12	&	-0.63	$	_{-0.44}^{+0.29}		$	&	0.18$_{-0.06}^{+0.05}$	\\
63	&	3.45$_{-0.02}^{+0.01}$	&	1.21	$\pm$	0.10	&	-1.57	$	_{-0.33}^{+0.24}	$	&	0.05$_{-0.01}^{+0.01}$	&	131	&	3.41$_{-0.02}^{+0.01}$	&	0.00	$\pm$	0.00	&	-2.88	$	_{-0.19}^{+0.16}		$	&	0.05$_{-0.00}^{+0.00}$	\\
64	&	3.46$_{-0.01}^{+0.04}$	&	3.46	$\pm$	0.16	&	-1.50	$	_{-0.27}^{+0.40}	$	&	0.06$_{-0.01}^{+0.06}$	&	132	&	3.46$_{-0.01}^{+0.04}$	&	0.00	$\pm$	0.00	&	-1.45	$	_{-0.19}^{+0.20}		$	&	0.08$_{-0.01}^{+0.06}$	\\
65	&	3.46$_{-0.01}^{+0.04}$	&	2.39	$\pm$	0.14	&	-1.47	$	_{-0.27}^{+0.40}	$	&	0.06$_{-0.01}^{+0.06}$	&	133	&	-	&	--	&	--	&	--	\\
66	&	3.46$_{-0.01}^{+0.04}$	&	1.99	$\pm$	0.14	&	-1.65	$	_{-0.27}^{+0.40}	$	&	0.07$_{-0.01}^{+0.05}$	&	134	&	3.58$_{-0.00}^{+0.00}$	&	--	&--	&	--	\\
67	&	3.46$_{-0.01}^{+0.04}$	&	6.26	$\pm$	0.22	&	-1.45	$	_{-0.27}^{+0.40}	$	&	0.06$_{-0.01}^{+0.06}$	&	135	&	3.58$_{-0.00}^{+0.00}$	&	--	& --	&	--	\\
68	&	3.42$_{-0.01}^{+0.02}$	&	4.27	$\pm$	0.35	&	-2.09	$	_{-0.31}^{+0.34}	$	&	0.02$_{-0.00}^{+0.01}$	&	136	&	3.50$_{-0.04}^{+0.03}$	&	4.40	$\pm$	0.00	&	-1.22	$	_{-0.44}^{+0.29}		$	&	0.18$_{-0.05}^{+0.06}$	
\end{longtable}
\end{spacing}

\normalsize

\twocolumn

\onecolumn

\begin{figure}
\label{fig:imagessed}
    \centering
   \includegraphics[width=1.0\textwidth,page=1]{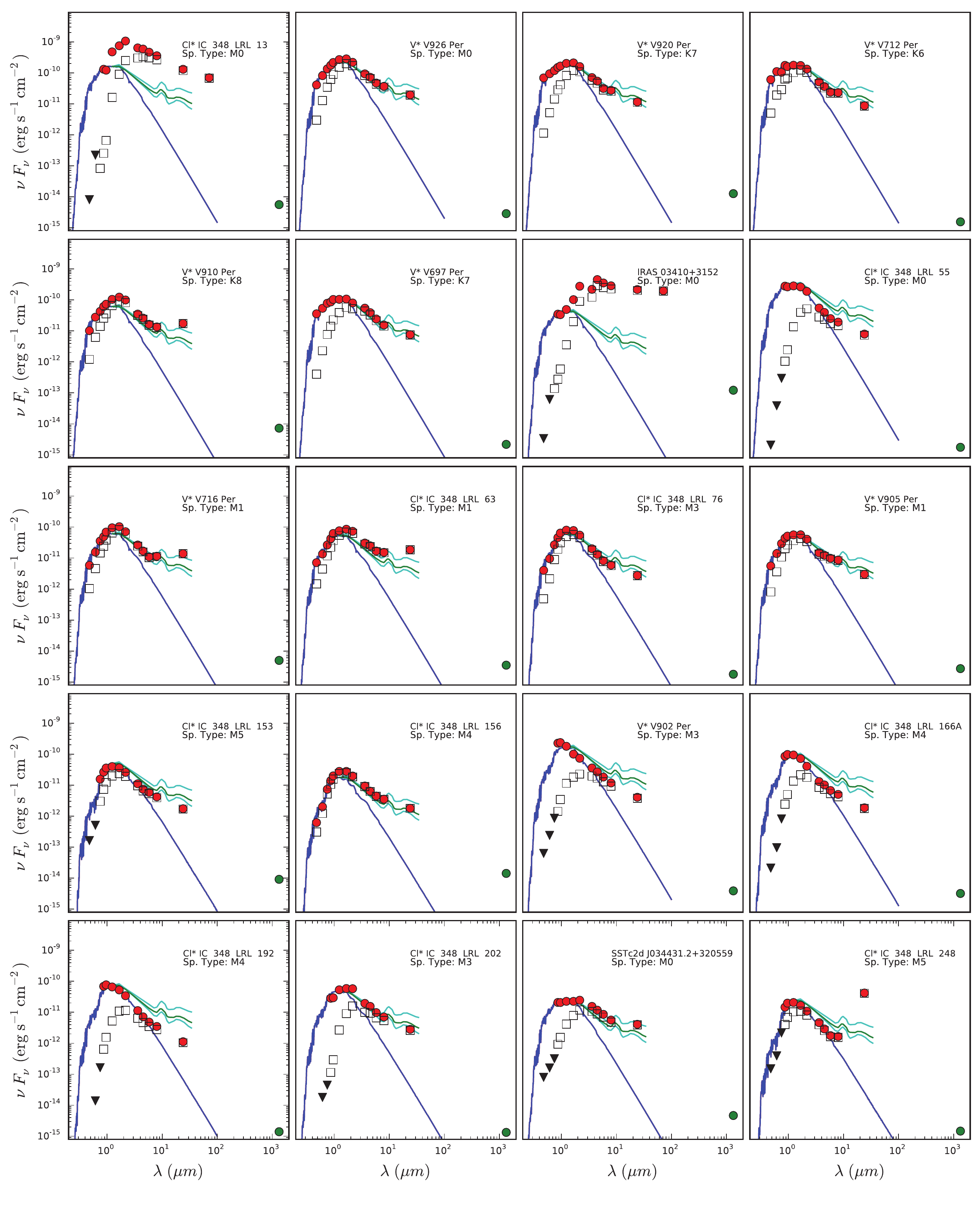}
    \caption[Spectral energy distribution of the sources in the IC~348 sample]{Spectral energy distributions of the sources detected at 1.3 mm in the IC~348 sample. Red dots show photometric data acquired from the literature; Green dots correspond to our ALMA integrated flux density values; blue lines are the BT-settl spectra model according to the spectral type. $A_v$ values used are in Table ~\ref{Table:properties}. The green lines correspond to the median SEDs of K5$-$M2 CTTSs calculated by \citet{Furlan2006}. The black boxes represent the observed optical and IR photometry
before correcting for extinction. Black triangles are optical photometry upper limits }
  \end{figure}

  \begin{figure}
  \label{fig:imagessed}
    \ContinuedFloat
    \captionsetup{list=off}
    \centering
    \includegraphics[width=1.0\textwidth,page=2]{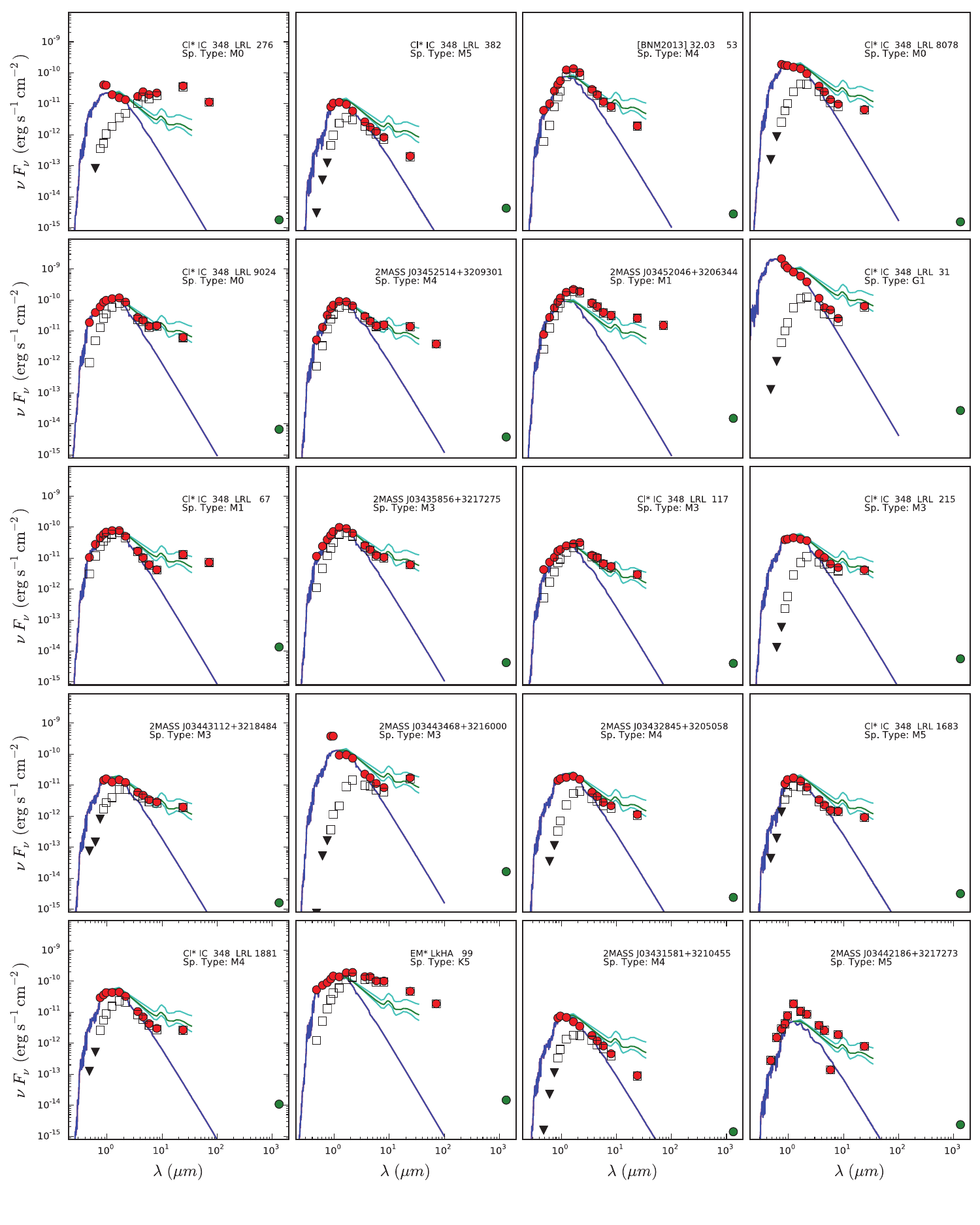}
    \caption{Continued.}
  \end{figure}

\twocolumn

 \begin{figure*}
 \centering
  \includegraphics[width=0.7\textwidth]{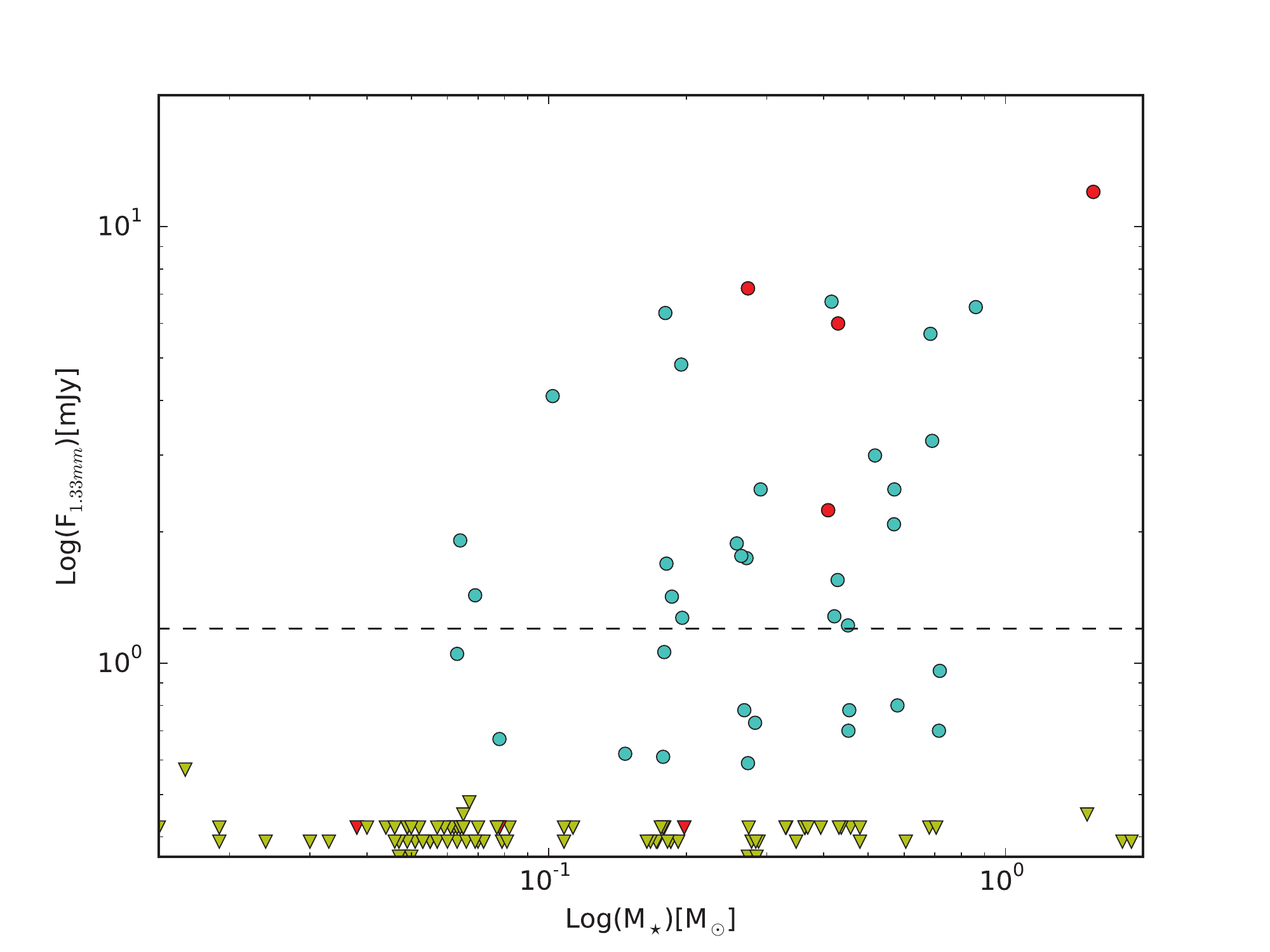}
   \caption[1.3 mm continuum flux as a function of stellar mass]{1.3 mm continuum flux as a function of stellar mass for IC~348 Class II sources. Cyan circles and yellow triangles represent detected sources and upper limits for non-detected sources, respectively. Four transitional disks (red circles) are among the most massive disks in the sample, while there are four transitional disks (red triangles) among the non-detected sources. The horizontal dashed-line indicates the flux level corresponding to a disk mass of $\sim$1 M$_{\rm Jup}$ assuming a gas to dust mass ratio of 100.}
  \label{fig:TDflux}
\end{figure*}

In Figure 5, we plot the SEDs of all targets detected at 1.3 mm, including photometry from PS1 (0.48, 0.62, 0.75, 0.87, 0.96 $\mu$m), 2MASS (1.25, 1.65, 2.22 $\mu$m) and Spitzer/IRAC (3.6, 4.5, 5.8, 8 and 24 $\mu$m)  \citep{Skrutskie2006, Evans2003, Currie2009}.  The photometric data were dereddened using the \citet{Mathis1990} approach. To calculate the stellar synthetic photometry with a fixed temperature T$_{\star }$, which is approximated by T$_{\rm eff}$, we interpolated the response curves for the set of filters used in the fitting, and used the BT-Settl spectral models for the corresponding T$_{\star }$ \citep{Allard2014}. Then, we convolved the filter response curves with the synthetic spectra, to match the spectral resolution. Because the PS1, 2MASS, IRAC, and 24 $\mu$m data have photometric uncertainties between a few percent and 0.1 mag for the objects investigated here, systematic effects can contribute up to 0.1 mag. To account for flux variability of the objects, we added an observational error of 15$\%$. A multiplicative dilution factor, $\left (\frac{R_{\star}}{d}  \right )^{2}$, relating the central star radius ($R_{\star}$) and the distance to the object ($d$) is used to normalize the optical bands.

In Figure~\ref{fig:TDflux}, we  plot millimeter flux as a function of stellar mass.  The 8 transition disks in our sample are indicated as red symbols.  Some of these objects are among the most massive disks in the cluster,  with disk masses of several M$_{\rm Jup}$,  assuming a standard gas to dust mass ratio of 100. In particular, 3 of the 6 brightest disks in the entire sample are transition objects based on their SEDs (Cl* IC 348 LRL 31, Cl* IC 348 LRL 67,  and 2MASS J03443468+3216000),  a trend that was already reported by \citet{Cieza2015} based on shallower SCUBA-2 observations of the cluster at 850 $\mu$m.

Three transition disks (Cl* IC 348 LRL 97*, Cl* IC 348 LRL 229*, Cl* IC 348 LRL 329) remained undetected. These results fit well in the scenario proposed by \citet{Owen2012} and \citet{Cieza2012}, in which there are at least two types of transition disks with inner opacity cavities that are the result of distinct processes:  1) gas-accreting transition disks that are massive and have large inner holes caused by the formation of giant planets,  (multiple) lower mass planets or subsequent migration \citep{Vandermarel2018}, and 2) non-accreting transition objects with low disk masses that have inner holes carved by photoevaporation during the final stages of disk dissipation.

\section{Discussion}
\label{Sec:Discussion}

\subsection{Non-Detections}

\begin{figure}
  \centering
     \includegraphics[width=0.5\textwidth]{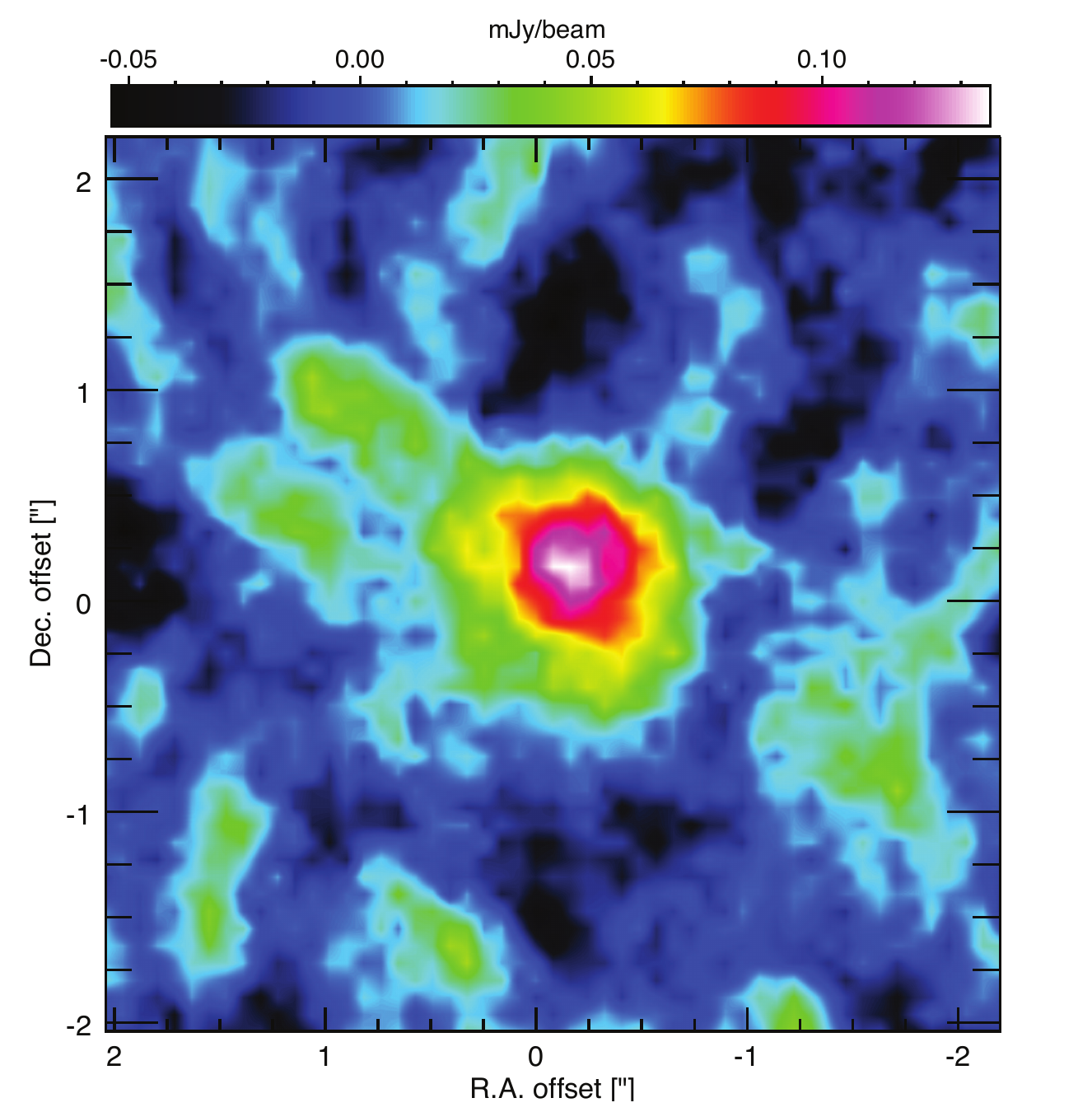} 
   \caption[Stacked image for the 96 non-detections]{Stacked image for the 96 non-detections, clearly showing a detection at the 6$\sigma$ level.}
 \label{Fig:stacked}
\end{figure}

The ensemble of undetected sources can be used to estimate the typical disk mass of the faint sources in IC 348. Initially, we stacked different spectral type sub-groups of these non-detections to constrain their average properties. We do not obtain significant detections in stacked images. Therefore, we stacked the 96 non-detections, after centering each field on the expected stellar position, to create an average image that has noise which is a factor of $\sim$7 lower than in the individual fields.  After doing so, we find a clear signal of 0.14 $\pm$ 0.02 mJy (Figure \ref{Fig:stacked}), indicating  that there are many targets in IC~348 with fluxes very close to the 1$\sigma$ noise of our observations.  
The 0.14 mJy flux measurement resulting from the stacking exercise suggests that the average dust mass of the disks that were not individually detected is only $\sim$0.40 M$_{\oplus}$.  
This implies that, for most disks in the IC~348 cluster, the amount of millimeter-sized dust that is still available for planet formation is of the order of the mass of the planet Mars. \emph{Kepler} has recently found that  M-type stars host an average of 2.2 $\pm$0.3 planets with radii of $\sim$1  R$_{\oplus}$ and orbital periods of 1.5 to 180 days \citep{Gaidos2016};

therefore, it is expected that most stars in IC~348 should form multiple rocky planets even though most of the cluster members have already lost their disks (within the stringent limits imposed by the infrared observations) or have very little dust left. Thus, we conclude that most disks around IC~348 members contain several Earth masses worth of solids in bodies that are at least several cm in size i. e., large enough to become undetectable by ALMA observations. More significantly, this suggests that these protoplanetary disks are likely sites of recently formed planetary systems like our own. In addition, IR emission from disks not detected at mm wavelengths connotes the existence of small, optically thick disks with extensions of $<$ 1 au. Our observations also constrain the amount of second-generation dust produced in the systems not detected by ALMA to be $<$ 0.4 M$_{\oplus}$, which still leaves significant room to explain the observed IR excesses. In fact, a small amount of warm grains of micron sizes (<1 lunar mass) is sufficient to produce the observed excesses at 10 $\micron$ \citep[e.g.][]{Nagel2010}.

In addition, our survey RMS of $\sim$0.15 mJy results in a large number of non-detections with respect to other surveys, mainly because of a lower sensitivity at late spectral types (M4-M9) at a distance of 310 pc. Detecting such late M stars individually (with S/N of $>$4) at 1.3 mm would require 10$\times$ our exposure time. However, we note that objects with a disk mass of $\sim$1 M$_{\oplus}$ are individually detected with a S/N of $\sim$10 in the Lupus survey thanks to the much smaller (150 pc) distance of some of the Lupus PMS stars and, to a lesser extent, the use of a shorter observing wavelength \citep{Ansdell2016}. This implies that at a distance of 150 pc and a sensitivity of 0.45 mJy in Band-6, it should possible to detect disks with dust masses of only $\sim$ 0.26 M$_{\oplus}$.

\subsection{Disk Evolution}

\begin{figure}
  \centering
     \includegraphics[width=0.5\textwidth]{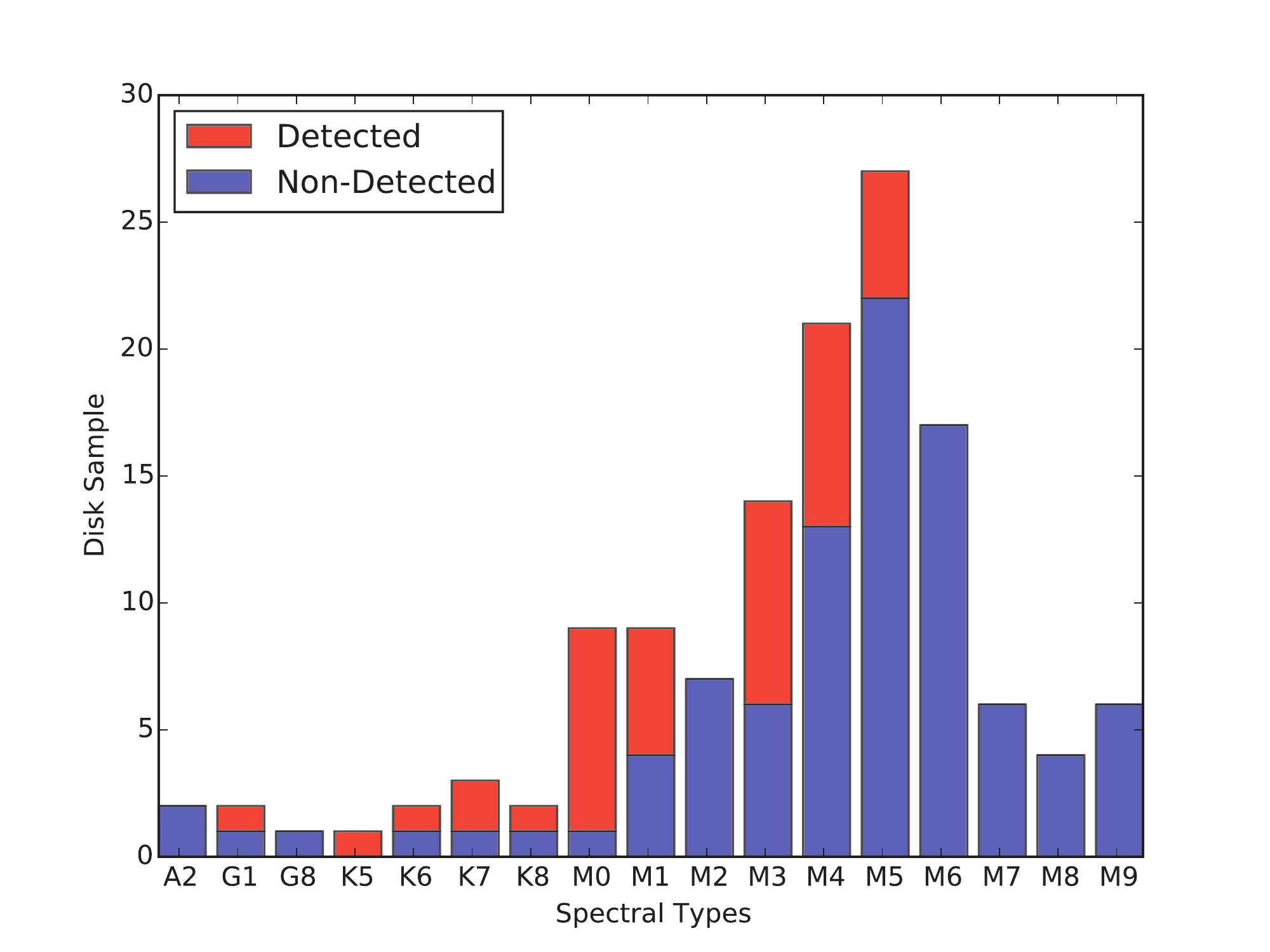} 
   \caption[Distribution of stellar spectral types for the detected and non-detected sources in IC 348 targeted]{Distribution of stellar spectral types for the detected and non-detected sources in IC 348 targeted by our ALMA survey (Tables \ref{Table:observations_no} and \ref{Table:observations}).}
    \label{Fig:detect_hist}
\end{figure}

Disk properties  determine possible planet formation scenarios. Investigating basic disk parameters such as mass and size at different evolutionary stages is thus vital for planet-formation theory. Nearby star-forming regions like Taurus (1-3 Myr), Lupus (1-3 Myr), Cha I (2-3 Myr), $\sigma$ Ori (3-5 Myr), and Upper Sco (5-10 Myr) are ideal targets to track evolutionary patterns because the ages of these populations cover the disk dispersal timescale. Recently, (sub-)mm continuum flux surveys of these star-forming regions have shown that disk masses decline with age and that there is a strong dependence of mm-wavelength luminosity on stellar mass \citep{Andrews2013, Ansdell2016, Ansdell2017, Barenfeld2016, Pascucci2016}.
Therefore, in order to compare IC 348 to other regions and investigate the evolution of disk masses as a function of stellar age, we need to take into account that disk masses and millimeter detection rates depend on spectral types and stellar mass. Figure \ref{Fig:detect_hist} displays the distribution of stellar spectral types for the detected and non-detected sources, showing the low detection rate at later spectral types (Tables \ref{Table:observations_no} and \ref{Table:observations}). 

\begin{figure*}
  \centering
     \includegraphics[width=1.0\textwidth]{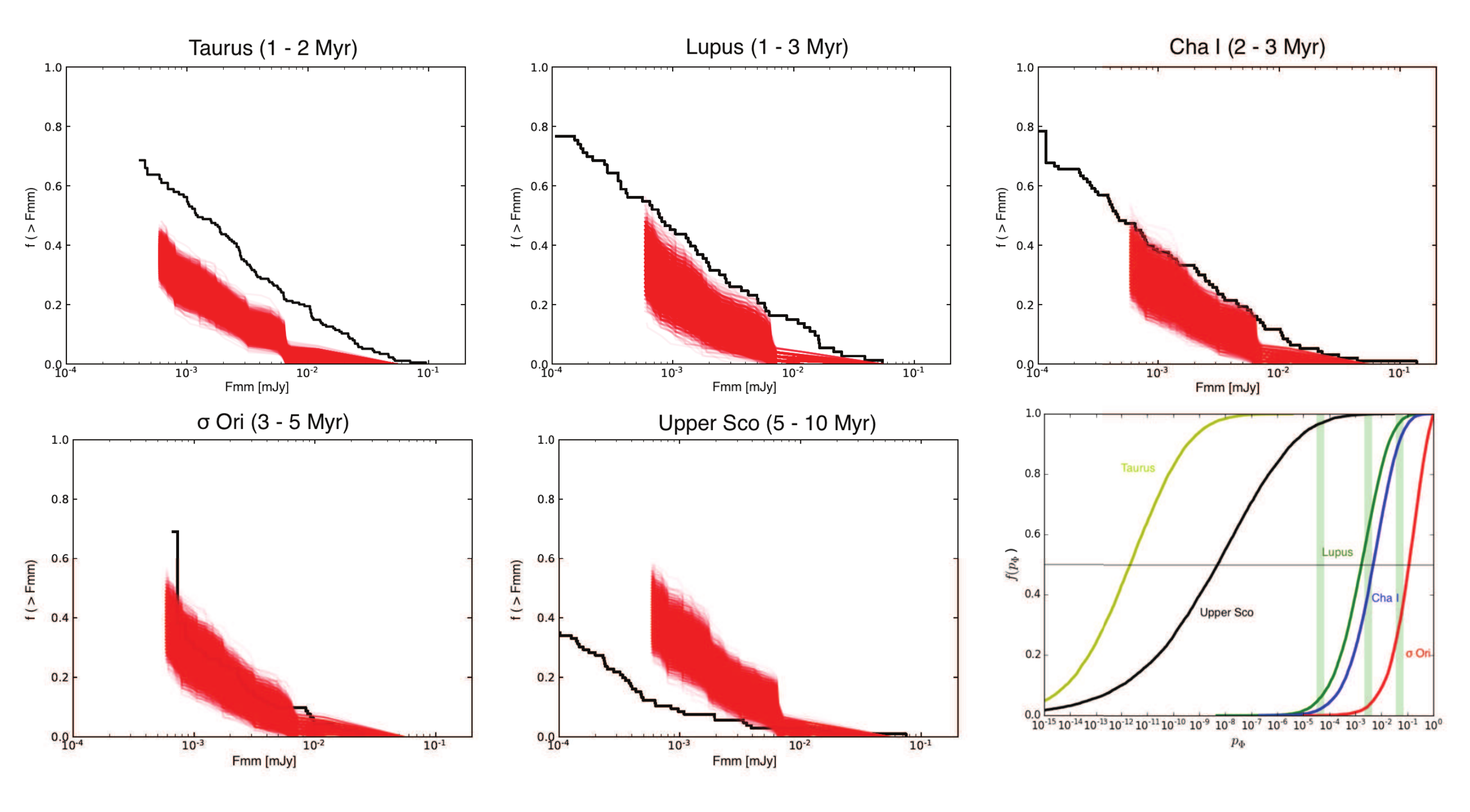} 
     \vspace{-0.2in}
   \caption[Cumulative Distribution Functions of the disk luminosities in IC 348]{Cumulative Distribution Functions of the disk luminosities in IC 348 (red) and the 10$^{6}$ synthetic ``reference'' disk draws, Taurus (1-2 Myr), Lupus (1-3 Myr), Cha I (2-3 Myr), $\sigma$ Ori (3-5 Myr), and Upper Sco (5-10 Myr), in black colour. At the right bottom, the comparison between the disk luminosity distribution of IC 348 and the ``reference'' sample shows the probability that IC 348 and the ``reference'' sample belong to the same population. The vertical green bars indicate the nominal 2, 3, and 4$\sigma$ probabilities.}
 \label{Fig:CDF}
\end{figure*}

Because estimates of stellar masses depend sensitively on inputs such as distances and theoretical models, here we used the statistical methodology presented in \citet{Andrews2013} based on spectral types.
While solar-mass stars evolve in spectral types during pre-main-sequence stages, lower-mass stars (0.1-0.7 M$_{\odot}$ evolve at almost constant temperature for the first $\sim$10 Myr (see evolutionary models in Figure  \ref{fig:baraffe}). This supports the use of spectral types as a proxy for stellar mass in the mass range of the stars in our IC~348 sample. 
Hence, to statistically compare samples from different regions, we perform  Monte Carlo simulations, whose ``reference'' sample is IC 348, while a ``comparison" sample can be Taurus, Chamaeleon I, Lupus, Upper Sco, or $\sigma$ Ori. The ``comparison"  sample is appropriately scaled to the IC 348 distance (310 pc) and modified for the respective observing wavelengths using the mean (sub-)millimeter flux ratios observed in Taurus (\rm F$_{\lambda}$ =  F$\rm _{1.3mm}$ $\times$ (\rm 1.3mm/$\lambda$)$^{2.5}$). Upper limit inputs for the ``comparison"  samples are as reported in the literature: three times the rms noise of the observations for Taurus, Lupus, Cha I, and Orionis, while the upper limits in Upper Sco are given by three times the rms noise plus any positive measured flux density. To construct our simulations, we first define a set of spectral type bins ranging from A2 to M6, corresponding to the distribution of the IC 348 sample, and place the comparison objects in those bins. Then, disk mm-wave luminosities are randomly drawn from  the reference region (IC~348) in each of these spectral type bins, such that the reference and comparison samples have the same spectral type distributions. In this manner, we simulate 10$^{6}$ synthetic ``reference'' disk ensembles that are used to construct Cumulative Distribution Functions (CDF); see Figure \ref{Fig:CDF}. Each of these CDFs is compared to the comparison sample to estimate the probability that the two distributions are drawn from the same parent population using a censored statistical test \citep[i.e the Gehan test:][]{Feigelson1985}. The result is a list of 10$^{6}$ such  probabilities for each comparison region. The cumulative distributions for these probabilities, $f(p_{\phi })$, are also shown in Figure \ref{Fig:CDF} (bottom-right panel).

\subsubsection{Relative Flux Densities}

The CDFs for the scaled flux densities show that the disks orbiting IC 348 stars are fainter on average than disks in Taurus, Lupus, Cham I and $\sigma$ Ori. \citet{Cieza2015} presented a similar statistical analysis based on shallower SCUBA-2 observations of IC~348  and found that the fluxes in this cluster were slightly lower than in Taurus. Here, we confirm that the fluxes in IC 348 are $\sim$ 4$\times$ fainter than in Taurus with a very high level of significance: virtually all 10$^{6}$ tests indicate that the probability that disk luminosities in Taurus and IC 348 are drawn from the same parent population is $<$ 10$^{-12}$. Similarly, we find that the younger Lupus region has a substantially brighter distribution compared to the older IC 348 region, at the $\gtrsim$3$\sigma$ level. The difference between the luminosity distributions of IC 348 and Cha I is marginal, with IC 348 being slightly fainter than Cha I, while, the luminosity distributions of $\sigma$ Ori and IC 348 are statistically indistinguishable ($\lesssim$2$\sigma$). In addition, Upper Sco is also very different from IC~348 (all tests indicate differences $>$ 3$\sigma$), but in the opposite sense: the Upper Sco disks are fainter than disks in IC~348, which reflects the fact that the mean dust mass is lower at the 5-10 Myr age of Upper Sco. In summary, these millimeter observations trace the population of millimeter/centimeter-sized grains at radial distances $>$10 au, confirming a significant dispersal process in the outer disk over a timescale of $\sim$1$-$10 Myr.

Infrared surveys with the \textit{Spitzer Space Telescope}, at IRAC wavelengths (3.6$-$4.5 $\micron$), previously established that the fraction of optically thick dust disk decreases with age, yielding disk fractions ($\%$) of 63 $\pm$ 4 in Taurus, 52 $\pm$ 5 in Lupus, 52 $\pm$ 6 in Cha I, 39 $\pm$ 6 in $\sigma$ Ori, 36 $\pm$ 3 in IC 348, and only 16 $\pm$ 6 in Upper Sco \citep{Ribas2014}. These IR observations probe the dispersion of micron-sized grains within a few au ($<$ 10 au) from the  central star. While IR disks observations are very sensitive and typically less biased with respect to spectral type, (sub-)millimetre detection rates are much lower and usually very biased against the lower end of the stellar mass function (M4-M9), making the interpretation of the results difficult.

\subsubsection{Continuum Luminosity Distributions}

Figure \ref{Fig:CDF} (bottom-right panel) compares the disc luminosity distributions of the ``comparison" and ``reference'' samples, where $p_{\phi}$ is the probability that the two distributions are drawn from the same parent population and the vertical green bars indicate the nominal 2$\sigma$, 3$\sigma$ and 4$\sigma$ probabilities. The cumulative distributions derived from the the Peto-Prentice test indicate medians of $p_{\phi}=$ 2.3 x 10$^{-12}$ and 5.5 x 10$^{-9}$ for Taurus and Upper Sco, respectively, implying a $>$4$\sigma$ difference. The Lupus and Cha I samples appear to have a difference of $\gtrsim$3$\sigma$ in their luminosity distributions, as indicated by medians of $p_{\phi}=$  1.7 x 10$^{-3}$ and $p_{\phi}=$  4.7 x 10$^{-3}$, respectively. Meanwhile, the $\sigma$ Ori has a luminosity distribution that is statiscally indistinguishable ($\lesssim$2$\sigma$) from the IC 348 sample, with $p_{\phi}=$ 1.1 x 10$^{-1}$.

It is noteworthy that the disc luminosity distribution of our IC 348 sample is significantly different from those of the Taurus and Upper Sco samples. As mentioned above, IC 348 is fainter than Taurus and Upper Sco is fainter than IC 348, which is not surprising, considering their relative ages and their IR disc fractions (Taurus: 63$\%$, Upper Sco: 16$\%$, IC 348: 36$\%$\citep{Ribas2014}). Also, $\sigma$ Ori, with an IR disc fraction of 39$\%$, seems to be at an evolutionary stage similar to that of IC 348 in terms of dispersal timescales.

\paragraph{Distance Uncertainty:}

Recent results for a similar analysis applied to the millimeter surveys of discs towards other star-forming regions (Taurus, Lupus, Cha I, Ori, Upper Sco), also reveal a significant decrease of disk masses with ``age'', which has been interpreted as a signature of evolution. However, ages are difficult to determine at early times ($<$10 Myr) and are highly dependent on the adopted distances and theoretical models \citep{Hillenbrand2008}. Here, we have adopted a ``representative" distance value  ($\sim$310 $\pm$20 pc) based on those IC 348 objects with high accuracy ${\it Gaia}$ DR2 parallaxes. However, our IC 348 sample presents a considerable dispersion in distance, even for objects with distance uncertainties $<$10$\%$. The closest object IC 348 12 is located at $\sim$214 $\pm$15 pc and the farthest object V* V697 Per is located at $\sim$352 $\pm$18 pc. If the distance is less than the adopted value of 310 pc, the target is expected to be even older than 5 Myr. For a distance of $\sim$260 pc, the luminosities would decrease by approximately 30$\%$ and the inferred age would be around 3-6 Myr \citep[e.g.][]{Ripepi2014}. Similarly, a larger distance would imply a younger age for a given target. If the distance is actually $\sim$360 pc, the luminosities would increase by approximately 30$\%$ and the mean age would be $\sim$1 Myr. Unfortunately, we do not have accurate ${\it Gaia}$ distance measurements for all IC 348 members studied here, but is its possible that not all targets are actual members of the cluster. Revising the membership status of the targets based on the ${\it Gaia}$ parallaxes and proper motions is beyond the scope of this paper, but vetting all regions for non-members would certainly be useful for future ALMA studies of disks in clusters. Here, our Montecarlo simulations are scaled to a distance of 310 pc and we emphasize that the foregoing comparisons of disc luminosity functions are highly dependent on the adopted distance and spectral types.

In addition, adopting spectral types as a proxy for mass also introduces uncertainties as pre-main-sequence stars, especially higher mass objects, can significantly  evolve in spectral type over time. Given that the mm-emission from disks depends on the host stellar mass, the results of our statistical analysis can be influenced by the difference in stellar masses at different evolutionary stages and spectral types.

\subsection{Disk mass vs. Stellar mass}
\label{Sec:linearfit}

\begin{figure}
  \centering
     \includegraphics[width=0.48\textwidth]{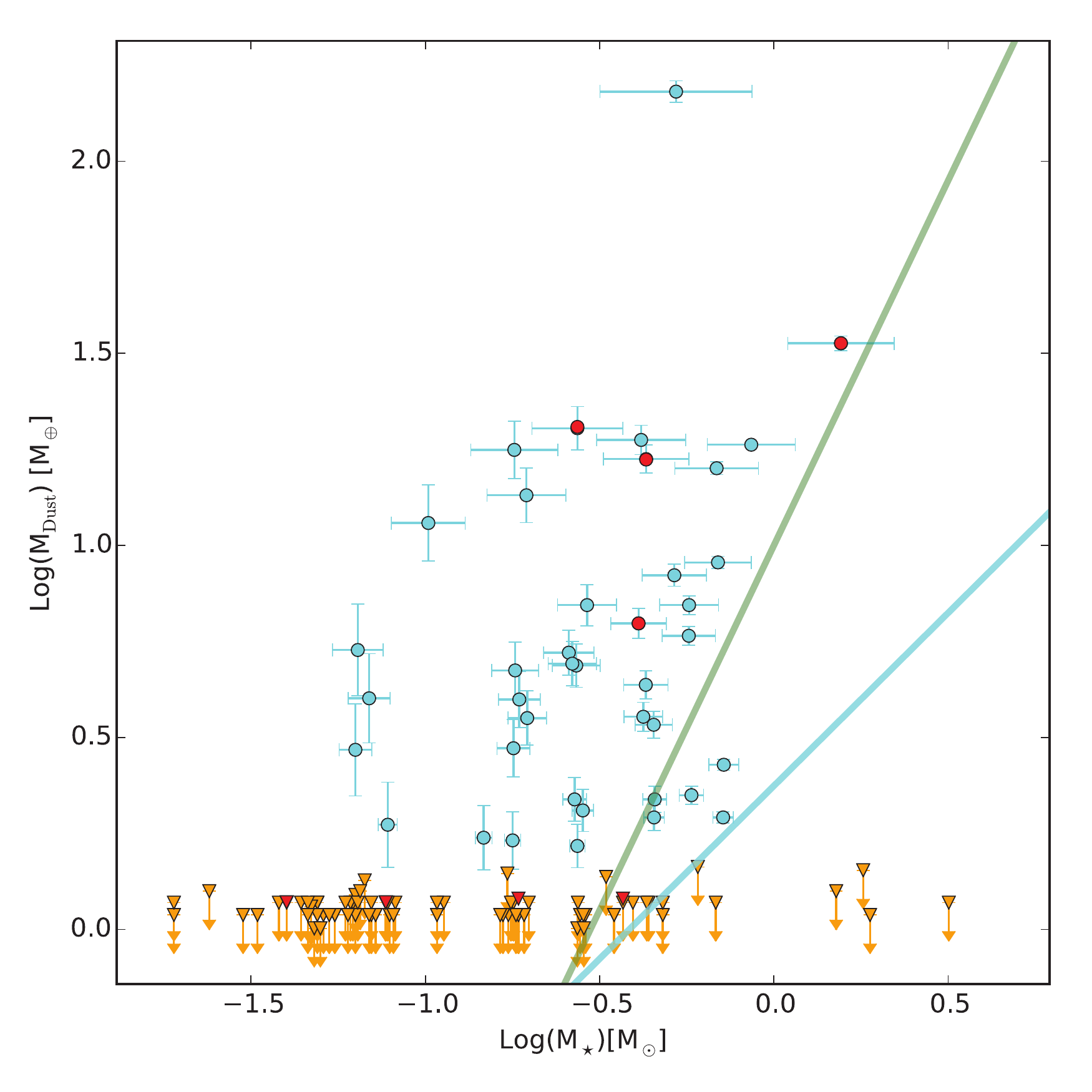} 
   \caption[Linear Fir]{Disk dust mass as a function of stellar mass for IC 348 region. Cyan circles represent the detected sources, while yellow triangles are 4$\sigma$ upper limits for non-detections. TDs are displayed by red circles and triangles, same as Figure \ref{fig:TDflux}. The cyan and green solid lines represent the Bayesian linear regression obtained for IC 348 and $\sigma$ Ori, respectively; see Section \ref{Sec:linearfit}. }
 \label{Fig:Linearfit}
\end{figure}

A commonly used approximation to estimate disk masses is the use of flux densities in the millimeter wavelength regime, where the disk luminosity is proportional to the dust mass \citep{Beckwith1990}. In recent years, a Bayesian linear regression approach analysis of ALMA surveys of star-forming regions at different ages have revealed a positive relationship between dust mass and stellar mass but with a steepening of the M$\rm _{dust}$ - M$\rm _{\ast}$ relation. \citep[e.g.][]{Andrews2013, Ansdell2016, Pascucci2016}. This method accounts for measurement errors in linear regression for detected and undetected sources, allowing one to correlate measurement errors, and to account for intrinsic scatter in the regression relationship \citep{Kelly2007}. Indeed,  studies of Taurus, Lupus, and Chamaeleon I show that the dependence of disk mass on stellar mass is similar at an age of $\sim$ 1-3 Myr, while older regions such as $\sigma$ Ori and Upper Sco present a steeper disk mass vs. stellar mass relation \citep{Pascucci2016, Ansdell2017}. The  steepening of this relation with age has been interpreted in terms of an efficient inward drift of mm-sized grains \citep{Pascucci2016}. However, the parameters describing the dependence of disk mass on stellar mass are very sensitive to the mm detection fraction and the treatment of the upper limits. In the case of IC 348, the detection fraction is low ($\sim$30 $\%$) and is a strong function of stellar mass. As a result, most of the detections are restricted to a narrow range of stellar masses. Given these issues, a linear regression fit is not accurate enough to allow a meaningful comparison to other star-forming regions. Nevertheless, we report the resulting parameters from the ``standard'' methodology used in the previous studies mentioned above. 

Considering all IC 348 sources in our ALMA sample, we derive slope and intercept values of $\beta = $ 0.90 $\pm$ 0.21 and $\alpha = $ 0.37 $\pm$ 0.18, where $\beta$ and $\alpha$  are the slope and intercept, respectively.  Figure \ref{Fig:Linearfit} shows the linear fit obtained from the Bayesian method. Because of the difficulty in obtaining a reliable fit, we only use $\sigma$ Ori as a comparison to illustrate differences between our fitting and other investigations with a wider mass range. The linear regression for  $\sigma$ Ori data generated values of 1.95 $\pm$ 0.37 and 1.00 $\pm$ 0.20 for $\beta$ and $\alpha$, respectively, and an intrinsic scatter value ($\delta$) of 0.65 $\pm$ 0.15, consistent with the values estimated by \citet{Ansdell2017}. The fitted linear regression for IC 348 provides a large intrinsic scatter of $\delta = $ 0.81 $\pm$ 0.15. Similar large intrinsic dispersions were estimated for Taurus, Lupus, Cha I, $\sigma$ Ori, and Upper Sco \citep{Pascucci2016, Ansdell2017}. As previously suggested by \citet{Pascucci2016}, the dispersion can be an intrinsic property of the disk population (i.e. disk masses, dust temperatures, and grain sizes) reflected in the diversity of planetary systems.

Moreover, M$\rm _{dust}$ measurements are subject to systematic uncertainties in the assumed parameters, such as distances to the star-forming regions. One can hence expect these results for M$\rm _{dust}$ to change with the availability of ${\it Gaia}$ DR2 data. In addition, obtaining a realistic dust temperature profile is important in the accuracy of estimates of M$\rm _{dust}$, and to this purpose, it is required a high-resolution data to generate those profiles. However, as shown by \citet{Tazzari2017}, assuming a constant T$_{dust}$ of $\sim$ 20 K, provides estimates of M$\rm _{dust}$ that are in good agreement with the results of more detailed modeling over a wide range of stellar masses,  0.1 to 2.0 M$_{\odot}$.

\subsection{CO Emission From IRAS 03410+3152}

The brightest millimeter source in our sample, IRAS 03410+3152, which has a bolometric temperature of 463 K and luminosity of 1.6 $\Lsun$ \citep{Hatchell2007}, was observed previously with the Submillimeter Array by \citet{Lee2011}. They detected a bipolar shape in the $^{12}$CO emission, with prominent emission outflow lobes and a moderate opening angle. From our observations at a resolution of 0.3$^{''}$, we are able to estimate position angle (P.A.), mass and kinematics of the of the outflow following the process presented in \citet{RuizRodriguez2017}. Here, we used the $^{13}$CO emission to correct for the CO optical depth and estimated the mass, momentum and kinetic energy of the outflow, see Figure \ref{Fig:12CO_13CO}. Using the C$^{18}$O line, we estimated a systemic velocity of $\sim$ 8.0 km s$^{-1}$. Because $^{12}$CO traces the bipolar and extension cavities of the outflow, we drew a line along the rotation axis to estimate a P.A. of $\sim$ -155$^{^{\circ}}$ north through east. Additionally, taking the extent of $\sim$ 3800 au (14$^{''}$) and maximum speed of the $^{12}$CO emission, we estimated a kinematic age of 1800 yr.

\begin{figure}
  \centering
     \includegraphics[width=0.45\textwidth]{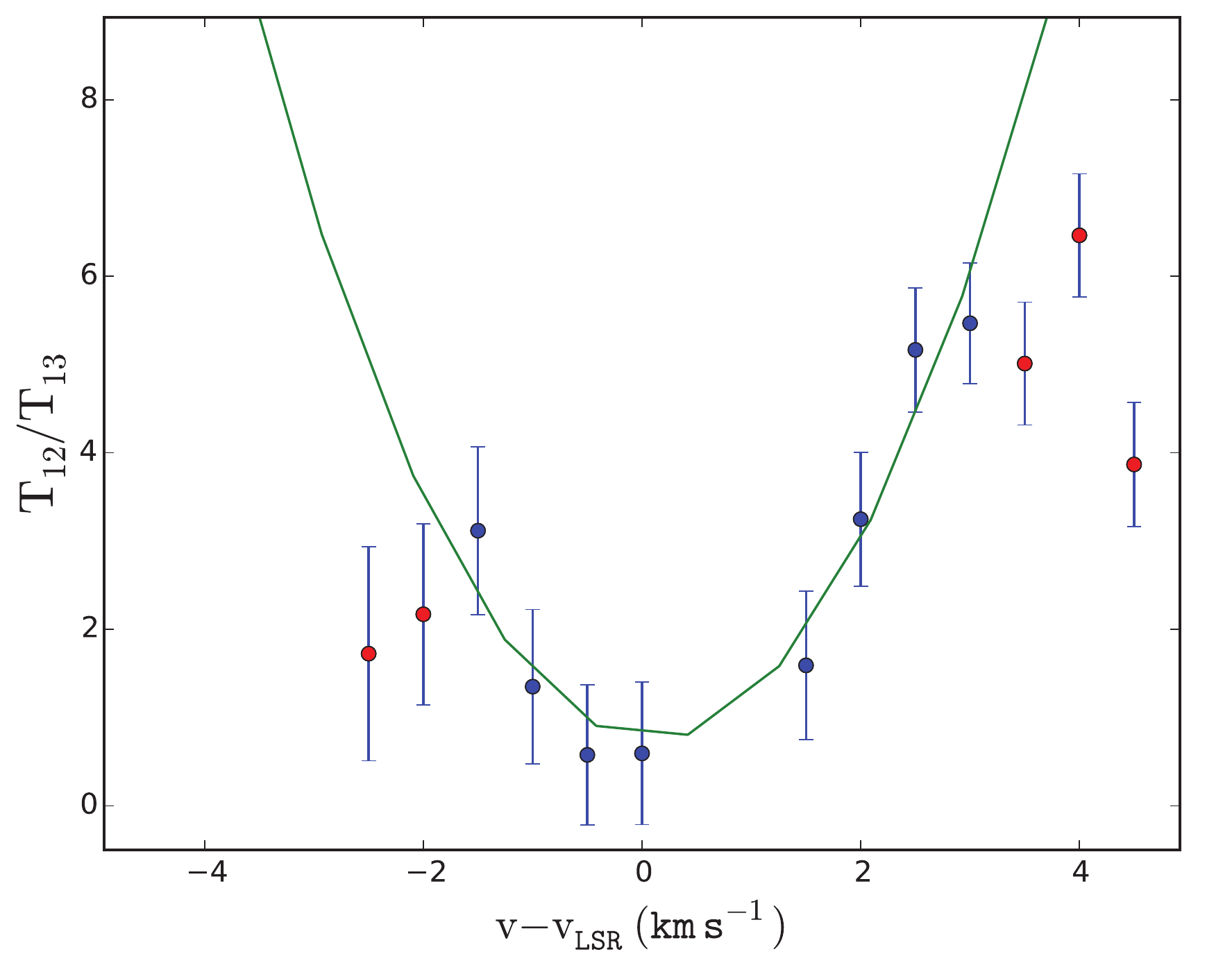} 
   \caption[Linear Fir]{Intensity ratio between $^{12}$CO and $^{13}$CO function of velocity. The green solid curve is the best-fit second-order polynomial using the blue data points, more details in \citet{RuizRodriguez2017}}
    \label{Fig:parabola}
\end{figure}

\begin{figure}
  \centering
     \includegraphics[width=0.5\textwidth]{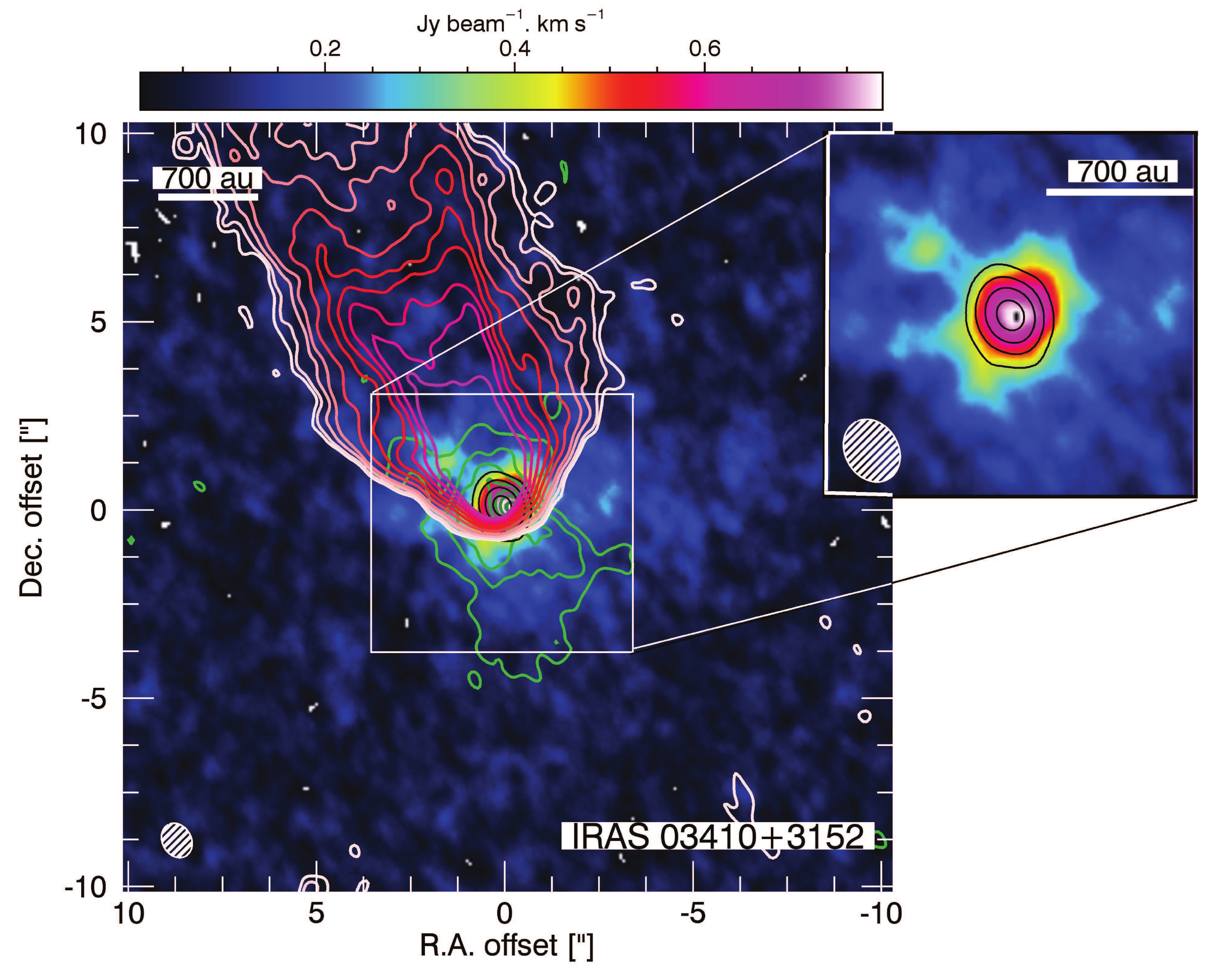} 
   \caption[IRAS 03410+3152]{$^{13}$CO intensity map (moment-0) of IRAS 03410+3152 integrated over the velocity
range of 5.0 to 11.5 km s$^{-1}$. Black contours show the 1.3 mm continuum emission around IRAS 03410+3152 at 10, 40, 60, and 80  $\times$ rms (0.15 mJy beam$^{-1}$). Green and Red contours show the blue- and red- shifted moment-0 of the $^{12}$CO line, respectively, at 20, 40, 80, 160 x 3$\sigma$ levels. These blue- and red- shifted intensity maps are integrated over the velocity range of 5.0 to 8 km s$^{-1}$ and 9.5 to 17.5 km s$^{-1}$, respectively. The synthesized beam of 0.77 arcsec $\times$ 0.63 arcsec with PA = 20.33 deg. is shown in the lower left corner. The upper right inset is a closeup ($\pm$2.5$^{''}$) of the central object.}

 \label{Fig:12CO_13CO}
\end{figure}

To compute the $^{12}$CO mass, we apply the correction factor to all the channels with $^{13}$CO detection above 4$\sigma$. In order to ensure emission only from the outflow, we built a mask around IRAS 03410+3152 of radius 3.0$^{''}$, where emission inside this area was removed from the integration. Thus, separating  the  red-  and  blue-shifted  components,  the  blue-shifted outflow kinematics were estimated by integrating channels in the range between 5.0 and 8.0 km s$^{-1}$ for $^{12}$CO and, 5.0 and 8.0 km s$^{-1}$ for $^{13}$CO. The range of channels in the redshifted emission is between 9.5 and 17.5 km s$^{-1}$ for $^{12}$CO and, 9.5 and 11.5 km s$^{-1}$ for $^{13}$CO. To apply the correction factor to all the channels with $^{12}$CO detection, we extrapolate values from a parabola fitted to the weighted mean values, where the minimum ratio value was fixed at zero velocity. In the fitting process, we did not include those data points presented as red dots in Figure \ref{Fig:parabola}, because at these velocities $^{12}$CO starts becoming optically thin. The fitted parabola has the form:

\begin{equation}
\centering
\frac{\textrm{T}_{12}}{\textrm{T}_{13}} = 0.11 + 0.40(\textrm{v-v}_{\textrm{\textsc{LSR}}})^{2}.
\label{eq:parabola}
\end{equation}

Table \ref{table:kinematics} shows the estimates at temperatures of 20 and 50 K and without correcting for inclination effects. Correcting for the $^{12}$CO optical depth increases the estimated mass of the outflow, the momentum and the kinetic energy by factors of 7.5-43, 5-27, and 4-23, respectively, at a temperature of 20 K.

IRAS 03410+3152 has been identified as an optically thick Class II protostar with a slope of $\alpha _{3.6-8.0\mu m}$ $\sim$ -0.006 \citep{Lada2006}. While the highly dereddened SED peaking in the mid-infrared clearly shows that IRAS 03410+3152 is still embedded, the presence of energetic outflows suggests that this object could be a Class I. Furthermore, the estimated outflow mass on the order of 10$^{-2}$ $\rm \Msun$ is consistent with the highest mass estimates of previously reported Class 0 and I outflows, after correcting for optical depth effects \citep{Dunham2014}. The differences between these estimates can be attributed to the higher ALMA sensitivities, which facilitate the detection of weak and high-velocity emission from the outflows, thus integrating over high-resolution spectra. In table \ref{table:kinematics}, note that the measured mass of the blue-shifted outflow is a factor of $\sim$2 lower than that of the red-shifted outflow, indicating possible differences in the environment between the cavities.

\begin{table*}
 \centering
 \begin{minipage}{128mm}
  \caption{Mass, Momentum, Luminosity and Kinetic Energy of the Outflow and Envelope}
 \label{table:kinematics}
  \begin{tabular}{clccccc}
 \hline \hline 
 \multicolumn{1}{c}{\textbf{}} &
  \multicolumn{1}{c}{\textbf{}} &
\multicolumn{2}{c}{\textbf{Red shifted}\footnotemark[1]}    &

\multicolumn{2}{c}{\textbf{Blue shifted}\footnotemark[2]}  \\[0.5ex] 

 \multicolumn{1}{c}{\textbf{Isotope}} &
 \multicolumn{1}{c}{\textbf{Property}} &
\multicolumn{1}{c}{\textbf{50 (K)} }  &
\multicolumn{1}{c}{\textbf{20 (K)}} &
\multicolumn{1}{c}{\textbf{50 (K)}}    &
\multicolumn{1}{c}{\textbf{20 (K)}}\\[0.5ex]\hline \hline \\[-3ex]  

\multirow{5}{*}{\begin{sideways}{\Large{ $^{12}$CO}} \end{sideways}}
&Mass  (10$^{-2}$ M$_{\odot}$)  &                                23.50 (163.40) & 15.89 (110.40)    &  1.68 (77.73)&   1.13 (52.53) \\ 
&Mass loss (10$^{-6}$ M$_{\odot}$ yr$^{-1}$)   &           131.18 (911.69)& 88.60 (616.14)  & 9.36 (433.80)& 6.33 (293.00)  \\ 
&Momentum  (10$^{-2}$ M$_{\odot}$ km s$^{-1}$)   &  106.00 (556.6)&  72.09 (376.21)&1.53 (40.95) &  1.04 (27.67) \\ 
&Energy (10$^{42}$ ergs.)   &                                        55.38 (220.39)& 37.43  (149.00) & 0.25 (5.50) & 0.16 (3.69)         \\ 
&Luminosity (10$^{-2}$L$_{\odot}$)   &                      25.46 (101.33) & 17.21 (68.48) & 0.11 (2.52)&  0.08 (1.70) \\[0.1ex]\hline

\multirow{5}{*}{\begin{sideways}{\Large{ $^{13}$CO}} \end{sideways}}
&Mass  (10$^{-2}$ M$_{\odot}$)  &  0.16 &0.11 & 0.08 & 0.55\\ 
&Mass loss ( 10$^{-6}$ M$_{\odot}$ yr$^{-1}$)  &  0.90   & 0.60    &4.63   & 3.10 \\ 
&Momentum  (10$^{-2}$ M$_{\odot}$ km s$^{-1}$) &    0.27  &0.18 &0.17&0.11\\ 
&Energy (10$^{41}$ ergs.)   &    0.49  & 0.33   & 0.24 & 0.16 \\ 
&Luminosity (10$^{-2}$ L$_{\odot}$)   & 0.02 & 0.02   & 0.01&  0.01  \\[0.1ex]\hline
\end{tabular}
$^{1}$ Blue shifted outflow kinematics were estimated after a cut above 4$\sigma$ and integration of channels  between 5.0 and 8.0  km~s$^{-1}$ for $^{12}$CO and 5.0 and 8.0  km~s$^{-1}$ for $^{13}$CO.\\
$^{2}$ Red shifted outflow kinematics were estimated with a threshold value above 4$\sigma$ and integration of channels between 9.5 and 17.5  km~s$^{-1}$ for $^{12}$CO, and 9.5 and 11.5  km~s$^{-1}$ for $^{13}$CO.\\
$^{3}$ Parameters inside the parentheses correspond to the computed values after applying the correction factors for optical depth effects to all the channels with $^{13}$CO detection above 4$\sigma$. 
\end{minipage}
\end{table*}

\section{Summary}
\label{Sec:Summary}

We have observed 136  Class II members of the young stellar cluster IC~348  with ALMA at 1.3 mm. We reach a dust mass sensitivity of 1.3 M$_{\oplus}$ (3$\sigma$) and detect a total of 40 disks.
The detection rate is a strong function of spectral type, as expected from the known dependence of disk mass on stellar mass. 
A stacking analysis of the  96 objects that were not individually detected yielded a clear 6$\sigma$ detection of 0.14 mJy, indicating that these disks have a typical dust mass of just $\lesssim$ 0.4 M$_{\oplus}$, even though their infrared SEDs remain optically thick and show little signs of evolution.  

We compare the disk luminosity function in IC~348 to those in younger and older regions and see a clear evolution in the dust masses between 1 and 5-10 Myr.
Based on the statistics of extrasolar planets \citep{Gaidos2016, Howard2012, Burke2015}, a stellar cluster like IC~348 with $\sim$400 members dominated by low-mass stars should form a very small fraction of systems ($\lesssim$5$\%$) with giant planets, which is consistent with the number of disks with masses $>$ 1 M$_{\rm Jup}$ in the cluster and the presence  of transition disks among this small population. 
The rest of the members should mostly form small rocky planets,  consuming most of the primordial dust by the age of the cluster. For the brightest millimeter source in our sample, IRAS 03410+3152, we are able to estimate P.A., mass and kinematics of the outflow. These estimates are characteristic of a Class-I type object.

\section*{Acknowledgments}

This paper makes use of the following ALMA data:
ADS/JAO.ALMA No. 2015.1.01037.S. ALMA is a partnership of
ESO (representing its member states), NSF (USA) and NINS (Japan),
together with NRC (Canada), NSC and ASIAA (Taiwan), and
KASI (Republic of Korea), in cooperation with the Republic of
Chile. The Joint ALMA Observatory is operated by ESO, AUI/NRAO
and NAOJ. The National Radio Astronomy Observatory is a facility
of the National Science Foundation operated under cooperative
agreement by Associated Universities, Inc.

D.R. acknowledges support from NASA Exoplanets program grant NNX16AB43G. L.A.C., acknowledges support from the Millennium Science Initiative (Chilean Ministry of Economy), through grant Nucleus  RC130007. L.A.C. was also supported by FONDECYT grant number 1171246. D. P. recognizes support by the National Aeronautics and Space Administration through Chandra Award Number GO6-17013A issued by the Chandra X-ray Observatory Center, which is operated by the Smithsonian Astrophysical Observatory for and on behalf of the National Aeronautics Space Administration under contract NAS8-03060. C.C. acknowledges support from project CONICYT PAI/Concurso Nacional Insercion en la Academia, convocatoria 2015, folio 79150049, and from ICM Nucleo Milenio de Formacion Planetaria, NPF. SC acknowledges support from FONDECYT grant 1171624.






\twocolumn
 \clearpage
\bibliographystyle{mn2e}
\bibliography{biblio}

\end{document}